\documentclass[12pt,preprint]{aastex}
\usepackage{natbib}
\usepackage{url}
\usepackage{amsmath}
\usepackage{appendix}
\usepackage{hyperref}
\begin{document}

\title{Regularly Spaced Infrared Peaks in the Dusty Spirals of Messier 100}

\author{Bruce G. Elmegreen\altaffilmark{1}, Debra Meloy Elmegreen\altaffilmark{2},
Yuri N. Efremov\altaffilmark{3}}

%altffilmark

\altaffiltext{1}{IBM Research Division, T.J. Watson Research Center, 1101 Kitchawan
Road, Yorktown Heights, NY 10598; bge@us.ibm.com}

\altaffiltext{2}{Department of Physics \& Astronomy, Vassar College, Poughkeepsie, NY
12604}

%\altaffiltext

\altaffiltext{3}{Sternberg Astronomical Institute of the Lomonosov Moscow State
University, Moscow 119992, Russia}

\begin{abstract}
Spitzer Space Telescope InfraRed Array Camera (IRAC) images of M100 show
numerous long filaments with regularly-spaced clumps, suggesting the associated
cloud complexes formed by large-scale gravitational instabilities in shocked and
accumulated gas. Optical images give no hint of this underlying regularity. The
typical spacing between near infrared (NIR) clumps is $\sim410$ pc, which is
$\sim3$ times the clump diameter, consistent with the fastest growing mode in a
filament of critical line density. The IRAC magnitudes and colors of several
hundred clumps are measured in the most obvious 27 filaments and elsewhere. The
clump colors suggest that the dust is associated with diffuse gas, PAH emission,
and local heating from star formation. Neighboring clumps on the same filament
have similar magnitudes. The existence of many clumps all along the filament
lengths suggests that the ages of the filaments are uniform. The observations
support a model where interstellar gas is systematically accumulated over
lengths exceeding several kpc, forming spiral-like filaments that spontaneously
collapse into giant clouds and stellar complexes. Optical wavelengths show
primarily the irregular dust debris, HII regions, and lingering star formation
downstream from these primal formation sites.
\end{abstract}

\section{Introduction}
\label{intro}

Local star formation is often in small filaments that form by transverse
compression and then gravitationally collapse in a semi-regular fashion into cores
\citep[e.g.,][]{andre10,miettinen18}. Sometimes these cores are at the
intersections of several filaments \citep{myers09}. On a galactic scale also,
stars form in filaments that may be shock fronts in spiral arms
\citep[e.g.][]{goodman14,ragan14} or rings from the expansion of superbubbles
\citep{egorov17}. Optical observations of galactic-scale star formation is
confusing, however, because extinction, irregular morphologies of star complexes,
and H$\alpha$ emission that is offset from the gas can make the intrinsic
regularities of filament collapse look more chaotic than it is.

\cite{baade} commented that in spiral galaxies, bright stars are ``strung out like
pearls along the arms.''  Early theoretical work considered shock compression
\citep{roberts69} and phase transitions \citep{shu72} as a way to enhance gas
self-gravity and star formation in the arms. \cite{mouschovias74} explained the
regular spacing of star formation with a magnetic Rayleigh-Taylor instability.
Star formation triggered by enhanced cloud collisions in the arms may also be
involved \citep{kwan83,scoville86}, and even cloud collisions can give a regular
spacing because of epicyclic motions \citep{dobbs08c}.

To study the regularity of spiral arm star formation in more detail,
\cite{elmegreen83} selected 22 galaxies with some evidence for regularity in the
optical images and measured the average ratio of the separations to the sizes of
the star-forming regions, obtaining a value of $3.1\pm1.2$. They suggested that
the regions formed by gravitational instabilities at the Jeans length for the
average gas density and velocity dispersion in the arms. Regularly spaced
star-forming clumps were observed more clearly in the spiral arm filaments of the
interacting galaxies NGC 2207 and IC 2163 \citep{elmegreen06}, where a
characteristic luminosity rather than the usual power-law luminosity function was
inferred for the regions in the brightest arm. \cite{efremov10} measured the
properties of regularly spaced star formation in M31, and \cite{gusev13} observed
it in NGC 628. Semi-regular star formation also occurs in nuclear starburst rings
\citep[e.g.,][]{pastoriza75,kennicutt89,elmegreen94a,crocker96,comeron10,laan13,vaisanen14},
molecular cloud cores \citep{keto91,sanchez14}, debris from interacting galaxies
\citep{wang04,bettoni10,tremblay14}, and galaxies along Mpc-scale cosmic filaments
\citep{tempel14}.

Theory suggests that regular fragmentation in filaments is an indication of
gravitational instabilities operating at the fastest-growing wavelength
\citep[e.g.,][see Sect. \ref{origin} below]{inutsuka92}.  Spiral arms filaments
presumably differ from local globular filaments \citep{schneider79}. Spiral arms
are often modelled as moving through gas over long distances, collecting it into
thin dust lanes in what may be a quasi steady-state \citep{roberts69}. Other
models suggest the filaments are transient \citep{dobbs08a,dobbs13}, building up
until they reach a point of instability, such as a critical line density, and then
collapsing into stars or dispersing by non-linear effects
\citep{chakrabartietal2003}.

The earliest theoretical models for gravitational collapse of gas in spiral arms
estimated the growth rates and flow-through times and compared the resulting
length and mass scales to the available cloud observations
\citep{elmegreen79,cowie81,tomisaka87}. Early simulations showed the build-up of a
magneto-gravitational instability in spiral arms \citep{kim01,kim02} and the
important role of gaseous self-gravity in forming giant molecular clouds
\citep{kim07}. Modern simulations show detailed cloud structure, including
regularly-spaced clumps in spiral arms with adaptive mesh gravitational
hydrodynamics \citep{renaud13}, and highly resolved cloud substructures using
particle hydrodynamics with phase transitions \citep{bonnell13}, molecule
formation \citep{dobbs08b,duarte16}, and star formation feedback \citep{dobbs11}.
Simulations also produce filamentary clouds, although in \cite{benincasa13} they
were irregular with typically one clump per filament, and in \cite{dobbs15b} and
\cite{duarte16} they were concentrated in the interarm regions as a result of
sheared spiral arm clouds. The observations in the present paper show both arm and
interarm filaments and in most cases they contain many clumps with a regularity in
their spacing and brightness that is not typically present in simulations
\citep[except, e.g.,][]{renaud14}.

Observations are still unclear about whether spiral arms trigger a net excess of
star formation in a galaxy or merely concentrate the gas in the arms, providing
the pearls-on-a-neckace appearance without changing the efficiency of star
formation per unit gas \citep{elmegreen86}.  There is abundant evidence that
molecular clouds and star-forming regions are larger in the arms
\citep[e.g.,][]{roberts87,colombo14}, but simulations that reproduce this effect
do not necessarily have an excess of star formation \citep{dobbs11,dobbs15a}, and
the large regions that form can be unbound and easily dispersed in the interarms
\citep{dobbs08c}. Observations by \cite{koda09} support this picture of loose
cloud agglomeration and interarm dispersal. A simulation by \cite{baba17} also
showed that molecular clouds go through the arms with little effect on their
internal properties and star formation. \cite{dobbs09} modelled cloud flow in a
spiral arm and reproduced well the Kennicutt-Schmidt relation between the surface
densities of star formation and gas, but they also got no enhancement in the
specific star formation rate in the arms.

Some observations show that molecular clouds are more self-gravitating in the arms
\citep{hirota11}, and others show they are not \citep{donovan13}. \cite{shabani18}
studied spiral arm triggering in a more conventional way, looking for age
gradients of star clusters across the arms \citep{yuan81}. They found cluster age
gradients in the symmetric two-arm spiral galaxy NGC 1566, but not in M51 or NGC
628, and suggested that M51 has a transient tidal arm rather than a steady wave
and NGC 628 is too weak to show a gradient. Another type of transient spiral was
modelled by \cite{baba13}. Transient spirals are not expected to have age
gradients because they can accrete gas from both sides \citep{dobbs10}.

A related observation is of spiral arm spurs or feathers, which can be regular too
\citep{sandage61,lynds70,elmegreen80,lavigne06,puerari14}. Spurs can arise when
the spiral arm condensations driven by self-gravity emerge into the interarms and
twist around in a locally reversed shear flow
\citep{balbus88,kim02,kim06,shetty06}. Shear and other instabilities may
contribute too, even without self-gravity
\citep{wada04,kim14,kim15,dobbs06,sormani17}. A comprehensive analysis of spur
formation with gravity and magnetic fields in a stationary two arm spiral was made
by \cite{lee12} and \cite{lee14}. \cite{renaud14} suggested that the regularly
spaced condensations inside spiral arms are from gravitational instabilities, and
the spurs that trail the arms are from Kelvin-Helmholtz-type instabilities.

To investigate the role of gravitational collapse in galactic-scale filaments, we
examined for this paper Spitzer Space Telescope InfraRed Array Camera (IRAC)
images of nearby spiral galaxies from the Spitzer image archive (e.g., NGC 300,
M74, M63, M83, M100, M101, NGC 6946, IC 342). All of these galaxies were found to
contain bright infrared (IR) condensations, or ``clumps,'' somewhat regularly
spaced along thin dust filaments that sometimes extend for several kpc. Because
these structures could be a key to understanding how spiral waves trigger star
formation, we chose one example, M100, and measured the clump properties from the
archival images. Clumps in the nuclear ring of M100 were studied previously by
\cite{knapen95}. The clumps in the spiral arms observed here are found to have
little correspondence with optical features, suggesting embedded or highly
obscured star formation if they are powered internally. This latter result needs
further study, perhaps with higher resolution IR obserations, because
\cite{prescott07} and \cite{schinnerer13} found relatively little embedded star
formation in the galaxies they observed. Our previous observations
\citep{elmegreen14} found a few regions that were visible in the IRAC bands and
invisible in the SDSS images.

The first IRAC observations showing clumpy spiral arm structure were for M81 in
\cite{willneretal2004}. \cite{calzetti05} studied clumps in the main spiral arms
of M51, combining $8\mu$m with $24\mu$m and other observations in $\sim500$ pc
apertures to determined star formation rates. \cite{foyle13} also observed
regularly spaced spiral arm clumps at IRAC $8\mu$m and with Spitzer and Herschel
FIR in M83; they found $10^6-10^8\;M_\odot$ gas masses on scales of 200-300 pc
with heating mostly by internal star formation. The structures reported in the
present paper are similar to these morphologically, but generally smaller and more
pervasive, with diameters of $\sim130$ pc and spacings of $\sim410$ pc.

\section{Observations}
\subsection{Morphology}

\begin{figure*}
\epsscale{1.}
%\plotone{f1.eps}
\plotone{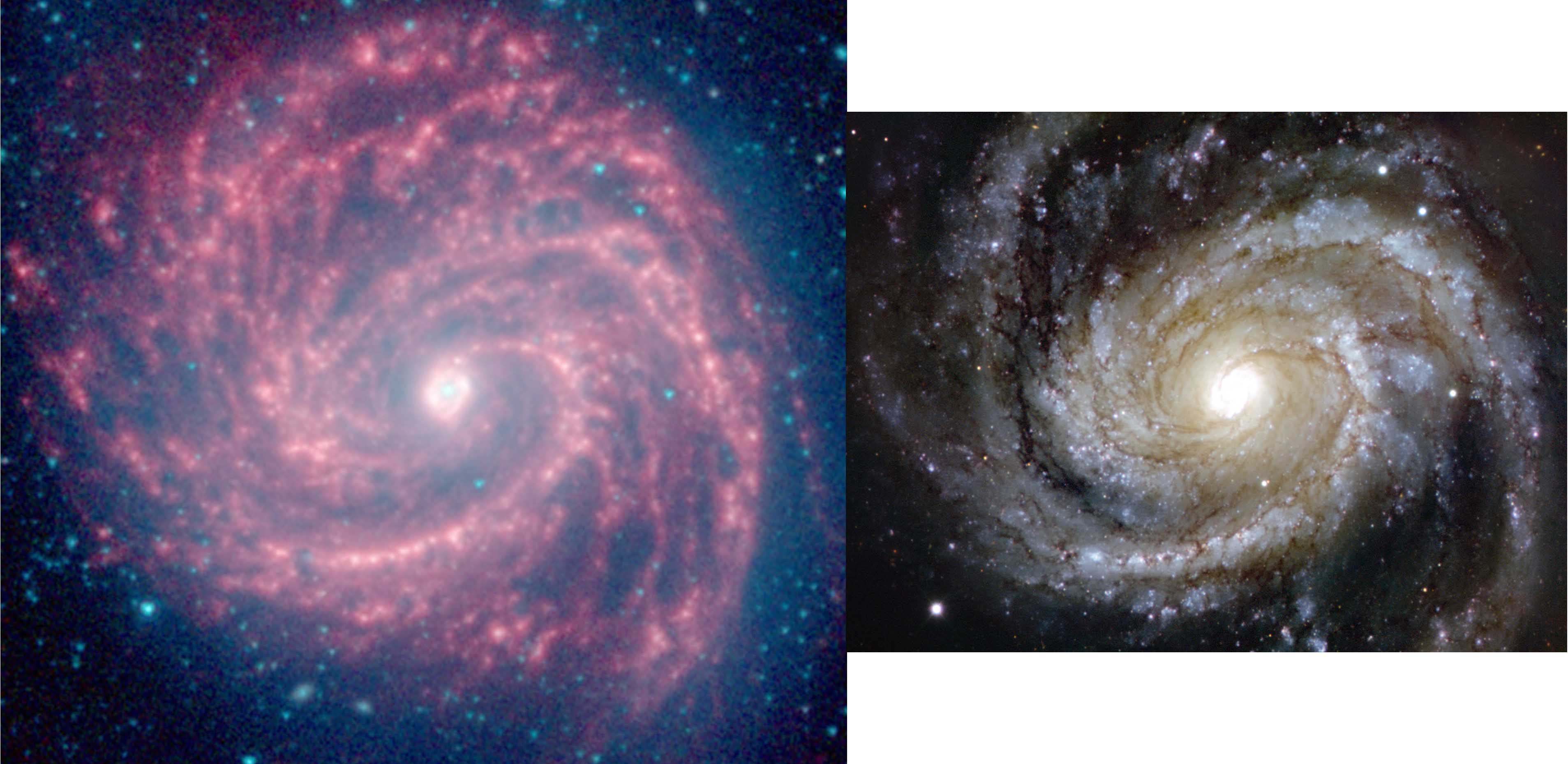}
\caption{Spitzer IRAC image of M100 at $3.6\mu$m, $4.5\mu$m, $5.8\mu$m, and $8\mu$m
on the left and VLT FORS image in optical bands R, V, and B to the same scale
on the right. North is up.  The Spitzer
image contains many filamentary regions ranging in scale from several kpc in the main
spiral arms to hundreds of pc in peripheral regions. The filaments typically contain
IR clumps that are spaced apart at approximately equal distances and have similar
magnitudes. The IR filaments and the regularity they reveal are not
evident in the optical bands, which highlight more the partially obscured HII regions
and irregular OB associations.}
\label{m100_ir_and_optical_rawforpub}
\end{figure*}

Figure \ref{m100_ir_and_optical_rawforpub} shows an IRAC image of M100 from the
Spitzer website\footnote{credit: NASA/JPL-Caltech; www.spitzer.caltech.edu/images/
5208-sig12-007-The-Swirling-Arms-of-the-M100-Galaxy }. The image is a composite of
$3.6\mu$m, $4.5\mu$m, $5.8\mu$m, and $8\mu$m images.  The galaxy is composed of
numerous clumpy filaments in dust emission, which make it different from the usual
view in optical images.  The right-hand side of Figure
\ref{m100_ir_and_optical_rawforpub} shows the Very Large Telescope (VLT) Focal
Reducer and Low-Dispersion Spectrograph (FORS) image of M100 in optical bands R,
V, and B to the same scale\footnote{credit: ESO;
http://eso.org/public/images/potw1330a/}. The thin filaments and semi-regular
spacings of clumps in the IRAC image are barely perceptible in the optical. Some
IR filaments are dust lanes in optical light without any evidence for the clumps,
some are haphazard strings of HII regions or bright irregular star complexes, and
some are completely indistinct.

To highlight this difference, Figure \ref{m100composite_clipped} shows
enlargements of select regions from the Spitzer and VLT images, displayed on the
same scale. Most of the IR clumps are regularly spaced and similar to each other
along the filaments, but the optical features are irregular in both extinction and
emission from H$\alpha$ and young stars.

Figure \ref{M100ch4shdivmips24shmult4_300bpi} shows the ratio of the $8\mu$m to
the Spitzer MIPS $24\mu$m emission. This ratio has the effect of removing the
local background to highlight small-scale NIR features, like an unsharp-mask
image. It occurs because the MIPS image has an angular resolution of
$7.1^{\prime\prime}$, the $8\mu$m image has a resolution of $2.4^{\prime\prime}$
\citep{chambers09}, and most of the $8\mu$m features have $24\mu$m counterparts.
The ratioed image brings out the clumps well, showing them as white dots
aligned along nearly all of the larger-scale structures.

\subsection{Measurement of Clump Properties}

The Spitzer image of M100 was examined for small bright clumps with central peaks
and red colors. Several small white clumps were avoided as they were thought to be
young regions that have partially broken through the dust. Stars are very blue in
these images and were easily avoided.

The position and flux of each clump were first measured in four IRAC bands using
the Image Reduction Analysis Facility (IRAF) task {\it phot} with a measurement
aperture of 2 pixels radius and background subtraction from an annulus between 3
and 4 pixels away from the center. This background region was chosen because the
clumps tend to be separated by 7 pixels (see below), and then the background lies
between the clumps.  The zero points for conversion of counts to magnitudes were
taken from the IRAC Instrument
handbook\footnote{http://irsa.ipac.caltech.edu/data/SPITZER/docs/irac/irac
instrumenthandbook/}. The positions and magnitudes for the clumps in the filaments
are given in Table 1.

Another set of measurements was made for the clumps on the filaments. Here, the
clump fluxes were determined with {\it phot} in a 1.5 pixel radius aperture with
no background annulus, and a background for subtraction was determined from other
${\it phot}$ 1.5 pixel apertures located midway between each pair of clumps. Thus,
for this second set, the background for subtraction was taken to be the average of
the fluxes from the two filament midpoints on either side of the clump; for clumps
at the ends of the filaments, the background was taken to be the flux from the one
adjacent filament midpoint.  This second measurement using filament midpoints as
backgrounds was designed to get the brightnesses of the clumps relative to the
adjacent filaments, and also the color excesses of the clumps relative to the
colors of the adjacent filaments. The clump magnitudes determined in this second
way are listed in Table 2, including only those with measurable interclump fluxes.

Clumps were selected for measurement based on their compactness and
brightness. For the first method discussed above, they are complete down to
$\sim13.5$ mag at $8\mu$m, $\sim15.5$ mag at $5.6\mu$m, $\sim18$ mag at $4.5\mu$m
and $\sim18$ mag at $3.6\mu$m. We determined these limits by blocking each chosen
clump with a black dot on Figure \ref{M100ch4shdivmips24shmult4_300bpi} as we
measured it, using the clump position to the nearest half pixel, and then
measuring fainter and fainter clumps until it was clear there were no remaining
small clumps brighter than the limits. This does not count multiple or highly
elliptical clumps or elongated bright regions that have larger total brightnesses,
as we are interested only in compact clumps like those that dot the filaments. We
also ran SExtractor to search for clumps on several types of images, including
Figures \ref{m100_ir_and_optical_rawforpub} and
\ref{M100ch4shdivmips24shmult4_300bpi}, but the background varies too much from
clump to clump to get any reasonable match to what we could find by eye.

In the end, we selected 422 clumps as representative of the whole galaxy. Of
these, 147 are in 27 filaments, most of which have a spiral-like appearance or are
in the main stellar density wave arms. Figure
\ref{m100_spitzer_with_map_try7_nostars_cropped} identifies the 147 filament
clumps using different symbols and colors for each filament. The 275 non-filament
clumps are shown as black dots. There are many more clumps that are diffuse or
faint that are ignored.

\subsection{Separation Distribution}

Figure \ref{efremovbeads_1} plots on the left the distribution function of the
separation between adjacent clumps on each filament with 3 or more clumps. The
units are in pixels on the Spitzer FITS images, which are $0.75^{\prime\prime}$ in
size. A pixel corresponds to 59 pc at a distance of 16.2 Mpc from
NED\footnote{https://ned.ipac.caltech.edu/}.  The peak in the distribution at 7
pixels corresponds to 410 pc.  This separation distribution does not consider the
inclination of the galaxy, which is $30^\circ$ \citep{knapen02}; deprojection
would increase the separations by at most 13\% along the major axes, which is not
significant.

The right-hand side of Figure \ref{efremovbeads_1} shows as a histogram the
relative difference in the separations between three adjacent clumps, calculated
as $2(S_{\rm i,i-1}-S_{\rm i+1,i})/(S_{\rm i,i-1}+S_{\rm i+1,i})$ for adjacent
clump indices $i-1$, $i$ and $i+1$ along a filament and separations $S$. The
relative difference in separation is small and it shows two peaks, one within 0.2
of 0 and another around 0.66.  The first peak corresponds to equal separations
between clumps, i.e., a regularity in their position along the filament. The
second peak corresponds to a gap in a regular spacing. That is, for a gap in the
midst of a regular spacing of 1 unit, the distance between the first two clumps
that straddle the gap is 2 units, and the distance between the next two clumps in
the sequence is 1 unit. Thus the relative difference is $2\times(2-1)/(2+1)=0.66$.
According to the right-hand side of Figure \ref{efremovbeads_1}, 48 relative
separations out of the total of 93 (52\%) in the plot are regular or regular with
a gap where there is one missing. This confirms the appearance by eye of the
regularity of the clumpy filaments in Figures \ref{m100_ir_and_optical_rawforpub}
to \ref{M100ch4shdivmips24shmult4_300bpi}.

Also on the right of Figure \ref{efremovbeads_1}, there are circles and horizontal
error bars representing the mean and variance, which show the number of clumps on
each filament ($y$-axis) versus the relative separations along these filaments
($x$-axis). Only filaments with more than 3 clumps are considered. The points show
a more precise regularity, i.e., a smaller number on the $x$-axis, for filaments
that contain fewer clumps. Longer filaments with many clumps have a slightly more
irregular spacing between the clumps. For clarity in plotting, the vertical
dimension is offset from the integer number of clumps in the filament by a small
random amount.

\subsection{Magnitude Distribution}

Figure \ref{efremovbeads_DF} shows, in blue, the apparent magnitude distribution
function at 8$\mu$m for all 422 clumps. This distribution is complete in the
bright part beyond the peak (see above). Figure \ref{efremovbeads_DF} also shows,
in red, the magnitude distribution for clumps in all of the identified filaments,
and in black, the clumps in the main spiral arm, which is to the west and south of
the center (the corresponding symbols in Figure
\ref{m100_spitzer_with_map_try7_nostars_cropped} are magenta squares, cyan squares
and green crosses). The main spiral arm has brighter clumps than elsewhere, as
evident also from the optical image. The luminosity scale at the top of this
figure is from the conversion of flux density into luminosity, assuming a width
for the IRAC4 filter equal to $2.88\mu$m, from the half-power points given in the
IRAC instrument handbook, page 18. The distance modulus is 31.05.

The right-hand part of Figure \ref{efremovbeads_DF} plots as a histogram the
differences in $8\mu$m magnitudes between adjacent clumps.  The circles and
horizontal lines are the means and variances of the magnitude differences for each
filament with more than 2 clumps (as opposed to more than 3 clumps for the
separation differences). The dispersion in the histogram for the differences is
0.46 mag., and the dispersion in the magnitudes themselves, from the left-hand
panel, is 1.01 mag. The expected dispersion of a random sample of differences
drawn from a Gaussian distribution like the magnitude distribution on the left is
$\sqrt{2}$ times the dispersion of the Gaussian, which would be 1.42 mag in our
case. The dispersion of the observed differences is only 0.32 times the expected
dispersion of the differences if the adjacent clumps were drawn from a random
sample. Thus the adjacent clumps are closer to each other in $8\mu$m brightness
than they would be in a random distribution, by a factor of 3.1. We get
approximately the same result considering only clumps brighter than 13.5 mag at
$8\mu$m for the total magnitude dispersion (0.76 mag in that case) and considering
only adjacent clumps where both are brighter than 13.5 mag at $8\mu$m (0.42 mag
dispersion for the mag differences), giving the ratio 2.5. These results, combined
with the regularity shown in Figure \ref{efremovbeads_1}, suggest that clump
formation in galactic-scale filaments is a coherent process.

\subsection{Color Distribution}

Figure \ref{efremovbeads_cmd_minusmid_combine_colordiff} shows color-magnitude and
color-color diagrams of the clumps, where the colors and magnitudes were
determined in the two ways mentioned above, once with background subtraction from
an annulus that lies between the clumps on average (top panels), and again with
the filament brightness used for the background subtraction (middle panels).  The
similarity of the $3.6\mu$m and $4.5\mu$m fluxes ([3.6]-[4.5] is small) and the
red colors at $5.8\mu$ and $8\mu$m indicate the presence of stars at the shorter
wavelengths and warm dust and PAH emission at the longer wavelengths.

In the top panels the $[3.6]-[4.5]$ color averages $0.20\pm0.03$ mag with
dispersion $\sigma=0.33$ mag; the $[4.5]-[5.6]$ color averages $2.17\pm0.03$ mag
with $\sigma=0.40$ mag, and the $[5.8]-[8.0]$ color averages $1.75\pm0.02$ mag
with $\sigma=0.19$ mag. In the middle panels, the $[3.6]-[4.5]$ color averages
$0.31\pm0.02$ mag with $\sigma=0.27$ mag; the $[4.5]-[5.6]$ color averages
$2.04\pm0.05$ mag with $\sigma=0.50$ mag, and the $[5.8]-[8.0]$ color averages
$1.71\pm0.02$ mag with $\sigma=0.21$ mag.

These colors compare well with models of emission from star-forming regions.
\cite{dale01} show model galaxy spectra with a jump equal to factor of $\sim5$
from $4.5\mu$m to $5.8\mu$m in their figure 5, and that factor corresponds to a
color of $[4.5]-[5.8]=1.75$ mag. \cite{gutermuth09} consider PAH colors as a
source of contamination for studies of protosars in star-forming regions; their
figure 15 shows the PAH region in color-color space where $[4.5]-[5.8]\sim1$ to 2
and $[3.6]-[4.5]\sim0$ to 1, as for our colors. Protostar envelopes are redder in
$[3.6]-[4.5]$ than our clumps because of their hotter dust, and protostar disks
are bluer in $[5.8]-[8.0]$ \citep{allen04}.  The IRAC colors of the diffuse
interstellar medium in the Milky Way are also close to our $[4.5]-[5.8]$ colors.
\cite{flagey06} measure these colors outside regions of star formation and
tabulate the ratios in their Table 1. Typically the ratio of $4.5\mu$m to
$5.8\mu$m emission is $\sim1/8$, which corresponds to a color of $\sim2.3$,
similar to that in Figure \ref{efremovbeads_cmd_minusmid_combine_colordiff}. The
ratio of $5.8\mu$m to $8.0\mu$m in \cite{flagey06} is smaller, $\sim0.3$, which
corresponds to a color of 1.2 mag, whereas we measure $[5.8]-[8.0]\sim1.5$ to 2 in
the top panels of Figure \ref{efremovbeads_cmd_minusmid_combine_colordiff}. This
difference implies that the $8\mu$m emission is relatively larger in the M100
clumps than in local diffuse clouds. Without longer wavelength observations at
comparable resolution (the clumps are a few arcseconds in size), we cannot make a
more complete spectral energy distribution (SED) and determine the contributions
from stars and dust, nor can we get the dust temperature and total dust
luminosity.

The bottom panels in Figure \ref{efremovbeads_cmd_minusmid_combine_colordiff} show
the excess colors of the clumps compared to the filaments. There is a lot of
scatter, but the clumps are slightly redder in $[3.6]-[4.5]$, by $0.044\pm0.005$
mag, than the adjacent filaments (with $\sigma=0.06$), about the same in
$[4.5]-[5.8]$ with an excess of only $0.042\pm0.010$ mag ($\sigma=0.12$), and
slightly bluer in $[5.8]-[8.0]$, with an excess of $-0.011\pm0.004$ mag
($\sigma=0.05$). These excesses suggest the filaments show a little more
underlying disk starlight than the clumps.

\subsection{Equivalent Stellar Masses}

The equivalent stellar masses that excite the clumps can be determined from the
IRAC fluxes and the bolometric magnitude of a young stellar population. These
stellar masses do not necessarily correspond to embedded stars because some of the
heating can come from adjacent stars that are visible optically. Nevertheless,
they provide a measure of the associated young stellar mass for each clump.

We first estimate the total IR luminosity for the clumps using the complete SEDs
of galaxies tabulated by \cite{xu01}. This tabulation gives flux density versus
wavelength for 6 normal galaxies with $24\mu$m luminosities in the range from
$10^6\;L_\odot$ to $10^{10.6}\;L_\odot$, and for 2 starburst galaxies with
$24\mu$m luminosities equal to $10^8\;L_\odot$ and $10^{11}\;L_\odot$. The SEDs
from \cite{xu01} were integrated over the total width of the four IRAC bands,
which is from $3.18\mu$m to $9.33\mu$m (the lower half-power point for the
$3.6\mu$m band and the upper half-power point for the $8\mu$m band, according to
the IRAC instrument handbook), and they were also integrated over the full SED to
give the total IR flux. The ratio of the total to the IRAC fluxes ranged from 22
for the lowest-mass normal galaxy to 12 for the highest mass normal galaxy, and it
was equal to 9.6 and 7.1 for the two starburst galaxies, respectively.  The high
values for low-mass normal galaxies reflect the dominance of background radiation
on the dust heating as there is little star formation in these systems. The values
decrease with increasing prominence of star formation because the PAH emission and
hot dust emission in the IRAC bands goes up relative to the cool dust emission at
longer wavelengths. We consider our clump SEDs to be most like the starburst SEDs
because they enclose or are adjacent to regions of active star formation, as
indicated by the juxtaposition of the clumps to HII regions and OB associations in
Figure \ref{m100composite_clipped}. Thus we take a ratio of $\sim8$ to convert the
summed flux in all four IRAC bands to the total IR luminosity.

This ratio is consistent with the ratios of total IR luminosity to $24\mu$m, which
is about 10, and $8\mu$m to $24\mu$m luminosity, which is about unity, for M51
where longer wavelengths were measured \citep{calzetti05}. It is similar also to
that in \cite{dale05}, who observe an approximately flat normalized flux density
distribution, $\nu f_{\nu}$ versus the logarithm of the wavelength for galaxies in
the SINGS survey, considering that the ratio of the log-wavelength interval for
the whole spectrum to the log-wavelength interval for IRAC is about 5 (that would
make the total IR flux $\sim5$ times the summed IRAC flux).

To convert the total IR luminosity to the mass of associated young stars, we use
the bolometric magnitude of a young stellar population given in \cite{bruzual03}.
This is $-2.7944$ for solar metallicity at less than 1 Myr age and $-1.0321$
at 10 Myr.  Using 4.74 mag as the bolometric magnitude of the Sun, the bolometric
luminosity per solar mass of young stars is
$3.01\times10^{35}\times10^{0.4*2.7944}$ at 1 Myr age, and the inferred stellar
mass is the inverse of this quantity multiplied by the total IR luminosity of the
clump.  All of the clump masses would be larger by the factor 5.1 for an age
of 10 Myr.

A histogram of the equivalent stellar masses for all 422 clumps at 1 Myr assumed
age is shown in Figure \ref{efremovbeads_mstaronly}. The average value of the log
of the mass, in solar masses, is $3.5\pm0.2$. This is reasonable for common, but
bright, star-forming regions in spiral galaxies. The sum of the masses of the
clumps in the filaments is $8.9\times10^5\;M_\odot$. If the typical clump age is
10 Myr, then the associated stellar mass increases to $4.5\times10^6\;M_\odot$ for
the filament clumps, based on the lower bolometric luminosity per unit stellar
mass given above.

\section{Origin of the IR Clumps}
\label{origin}

The equal spacings and similar magnitudes of the clumps on the filaments suggests
that the condensations formed by a regular process such as a gravitational
instability along the filament lengths.  The separation should be the length of
the fastest growing unstable mode. The appearance of multiple clumps on filaments,
sometimes with a half-dozen or more clumps, also implies that all parts of the
filament formed at about the same time. Then all of the clumps collapsed together
before the filaments could be dispersed by shear and star-formation feedback.

Filament instabilities without a magnetic field were studied early on by
\cite{ostriker64} and \cite{inutsuka92}, and with a magnetic field by
\cite{chandra53}, \cite{stod63}, \cite{nagasawa87}, \cite{nakamura93},
\cite{tomisaka95}, \cite{fiege00} and others. When the filament has the
equilibrium mass per unit length, $\mu=2\sigma^2/G$ for velocity dispersion
$\sigma$, these authors found a dominant wavelength, or separation between
condensations, that is about 3.9 times the effective filament diameter, $D_{\rm
eff}$, which is $D_{\rm eff}=2\mu(\pi\rho_{\rm c})^{-1/2}$ for central density
$\rho_{\rm c}$. The growth rate for this mode is $\omega=0.34\times(4\pi
G\rho_{\rm c})^{1/2}$. A filament confined by high pressure has a longer dominate
wavelength and a slower growth rate because the mass per unit length is less at
the same central density. A magnetic filament with an aligned field has about the
same dominant wavelength if it is not highly confined by pressure because the
instability occurs in a direction where the field exerts little force. If it is
confined by pressure, then the field slows the growth as the wavelength increases.

\begin{figure*}
\epsscale{1.}
%\plotone{f2.eps}
\plotone{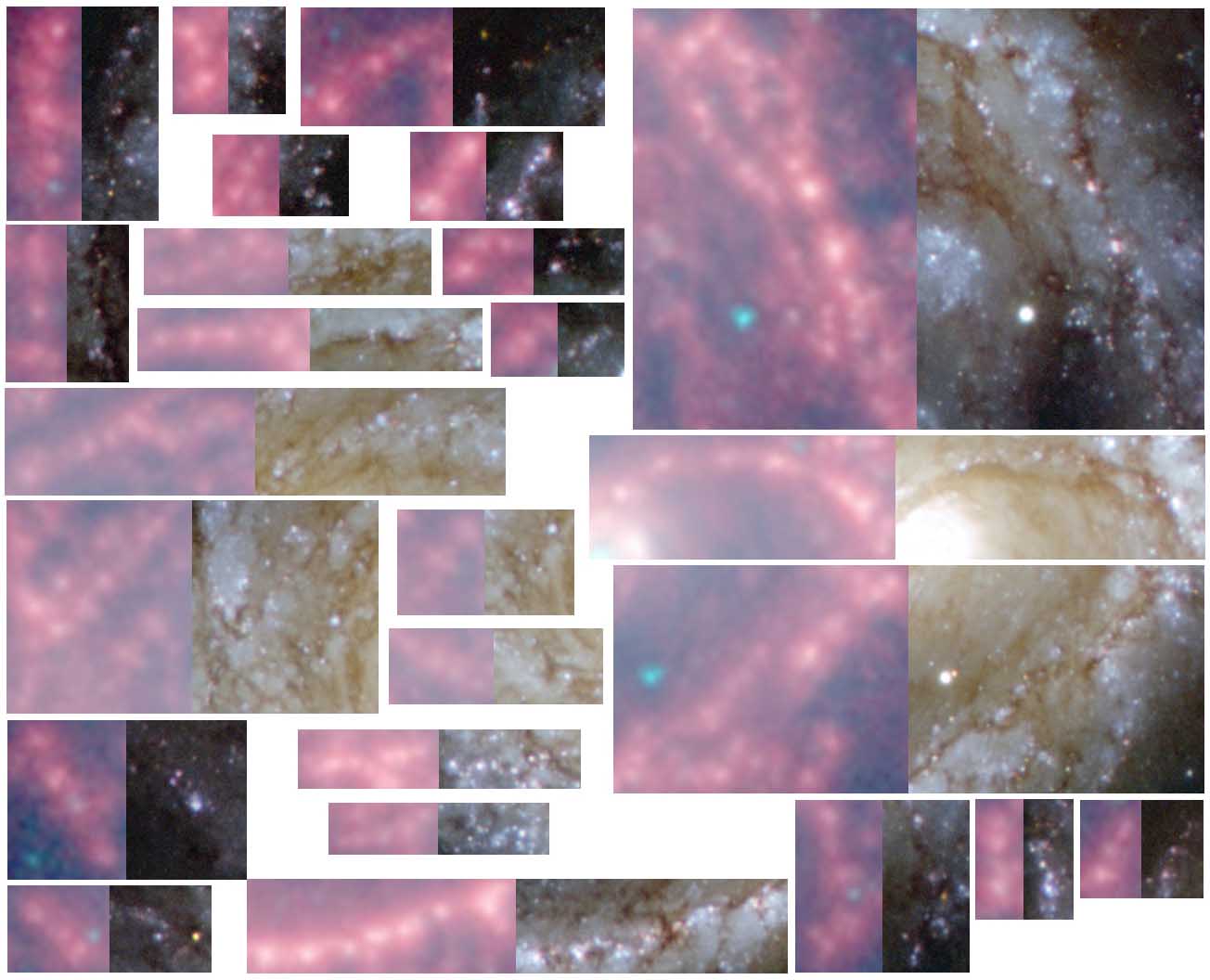}
\caption{Cut-outs from Figure \ref{m100_ir_and_optical_rawforpub} showing clumpy filaments from the
Spitzer IR image on the left and the same regions from the VLT optical
image on the right. The regularity of the star-forming clumps is clear in the
IR images, yet there is often little trace of it in the optical.}
\label{m100composite_clipped}
\end{figure*}

\begin{figure*}
\epsscale{1.}
%\plotone{f3.eps}
\plotone{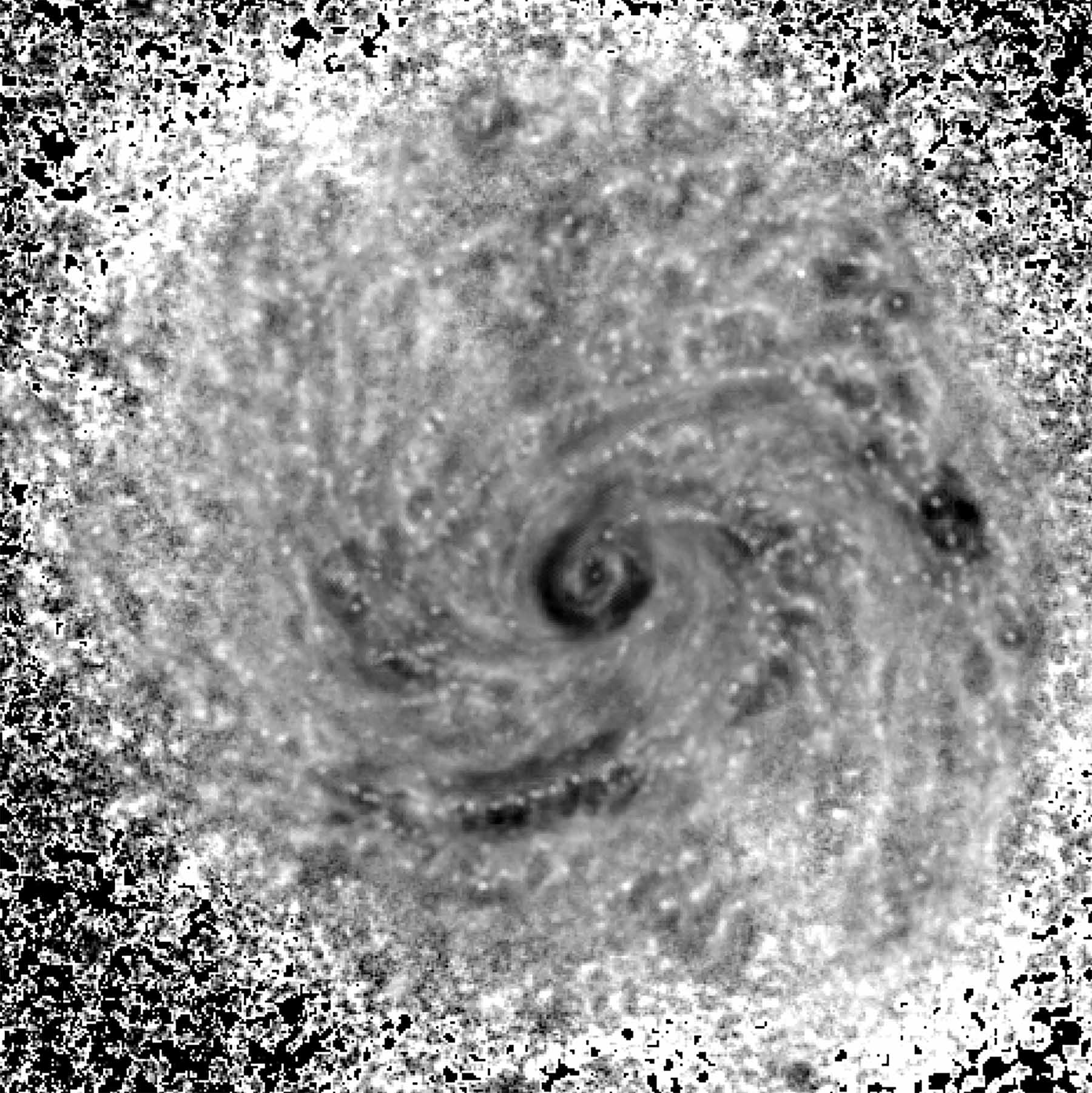}
\caption{IRAC $8\mu$m image divided by the MIPS $24\mu$m image of M100, plotted on a
log scale with the MIPS image multiplied by 4 to match the intensities. The numerous white
dots that line up on most of the spiral-like filaments and spurs show the IR
clumps studied here.  MIPS $24\mu$m emission also comes from these clumps, but the
angular resolution at $24\mu$m is several times larger than the clump size, so the MIPS image
provides a good model to divide out the large-scale structure and highlight the $8\mu$m clumps.
The dark regions around some concentrations of clumps are from the division by
broad $24\mu$m emission.}
\label{M100ch4shdivmips24shmult4_300bpi}
\end{figure*}

\begin{figure*}
\epsscale{1.}
%\plotone{f4.eps}
\plotone{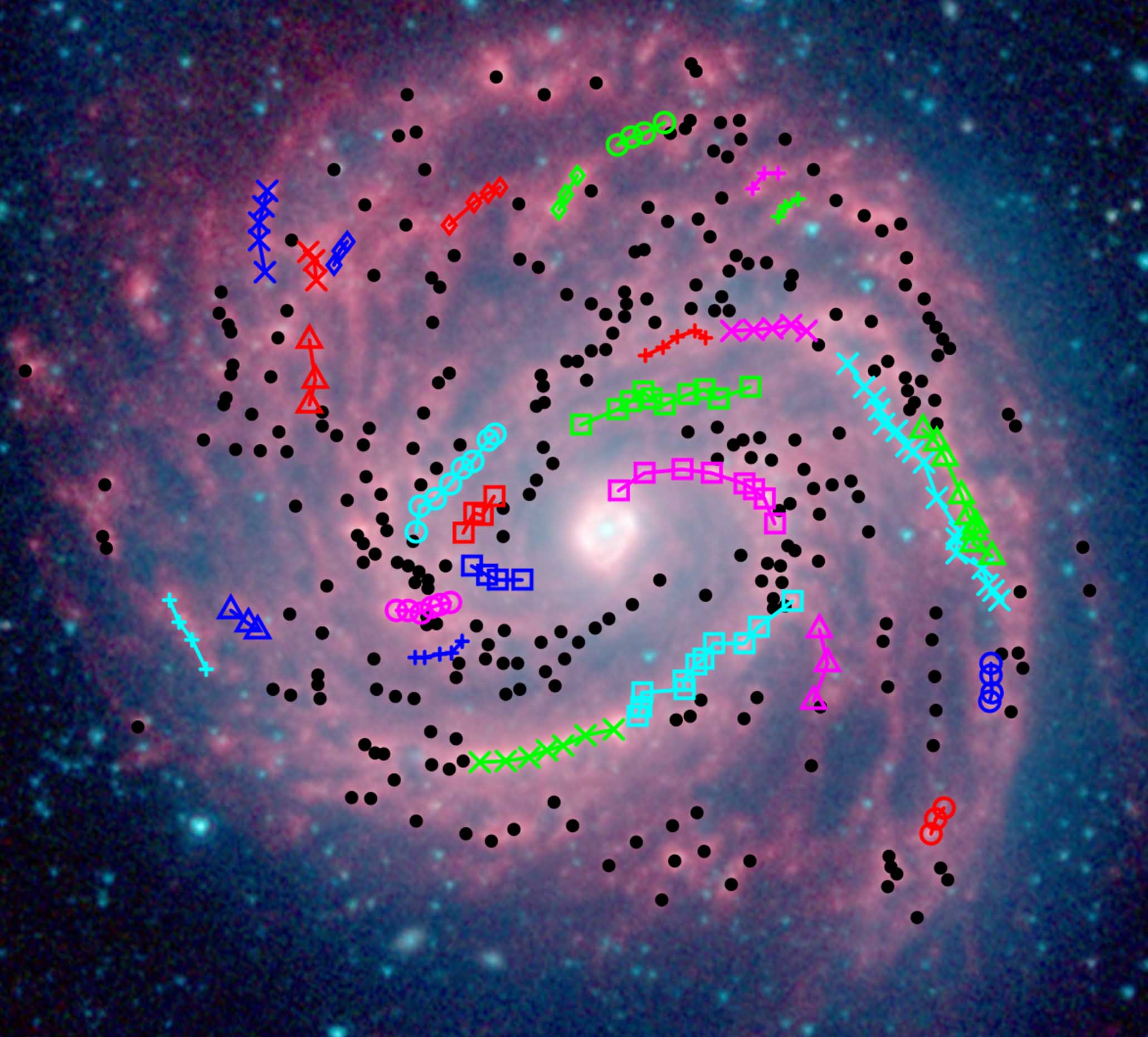}
\caption{The Spitzer image of M100 as in Figure \ref{m100_ir_and_optical_rawforpub}
with 422 bright clumps indicated,
including 147 with various symbols and colors assigned to 27 distinct filaments, as listed
in Tables 1 and 2.}
\label{m100_spitzer_with_map_try7_nostars_cropped}
\end{figure*}

\begin{figure*}
\epsscale{1.}
%\plotone{f5.eps}
\plotone{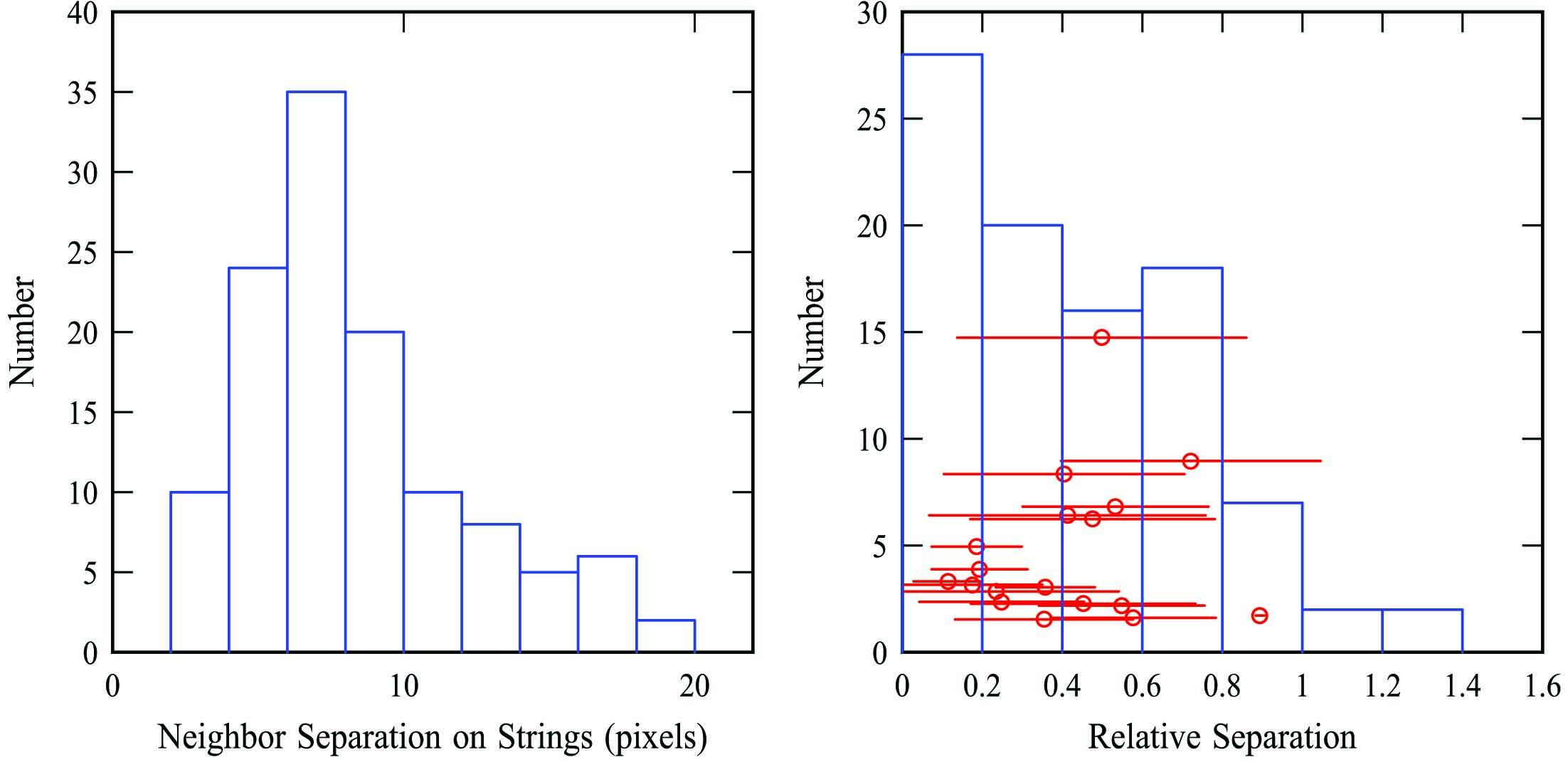}
\caption{(left) Histogram showing the distribution of 120 separations between
clumps on the 27 filaments indicated in Figure \ref{m100_spitzer_with_map_try7_nostars_cropped}.
The separations are
measured in pixels on the Spitzer image, which are $0.75^{\prime\prime}$ and
correspond to 60 pc. The peak occurs at 410 pc. (right) Histogram of the relative
separations between clumps on the filaments with 3 or more clumps. The relative
separation is the difference between two adjacent separations divided by their
average separation. The histogram peaks between 0 and 0.2 relative separations, which
corresponds to equally spaced clumps, and again between 0.6 and 0.8, which
corresponds to a gap between nearby equally-spaced clumps.
The circles and horizontal lines are the means and variances of the relative separations
for the individual filaments with more than 3 clumps.  The circle at (0.89,2) corresponds
to filament 17 with red squares in Figure 4; it has 4 clumps with the inner two close
together and the outer two separated by almost the same distance from their adjacent inner
clumps, giving a very small variance in the separation differences for each set of 3
contiguous clumps.}
\label{efremovbeads_1}
\end{figure*}

\begin{figure*}
\epsscale{1.}
%\plotone{f6.eps}
\plotone{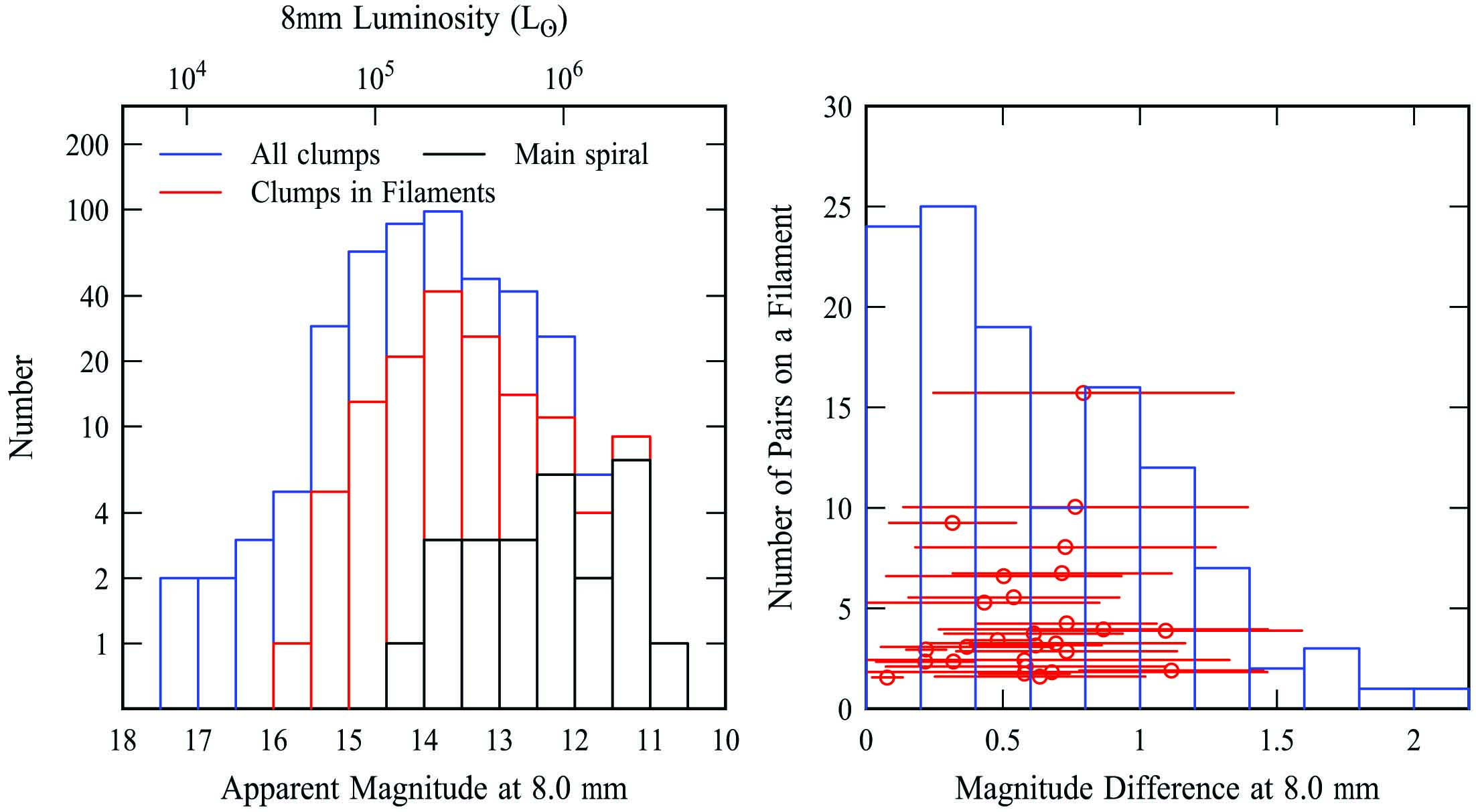}
\caption{(left) Histogram of the apparent magnitude at $8\mu$m of all 422 clumps
(in blue), all of the 147 clumps in the 27 filaments (red) and the 26 clumps in the
main spiral arms (black). The sample is complete to the right of the peak. The luminosity in the $8\mu$m
band is shown on the top axis. (right) The distribution of magnitude differences between
near-neighbor clumps on filaments. Their average difference in magnitudes is only one-third
of that expected from randomly sampling the filament distribution on the left, which means
that neighboring clumps are more like each other than a random pairing.
The circles and horizontal lines are the means and variances of the relative magnitude
differences for individual filaments with more than 2 clumps.
}
\label{efremovbeads_DF}
\end{figure*}

\begin{figure*}
\epsscale{1.}
%\plotone{f7.eps}
\plotone{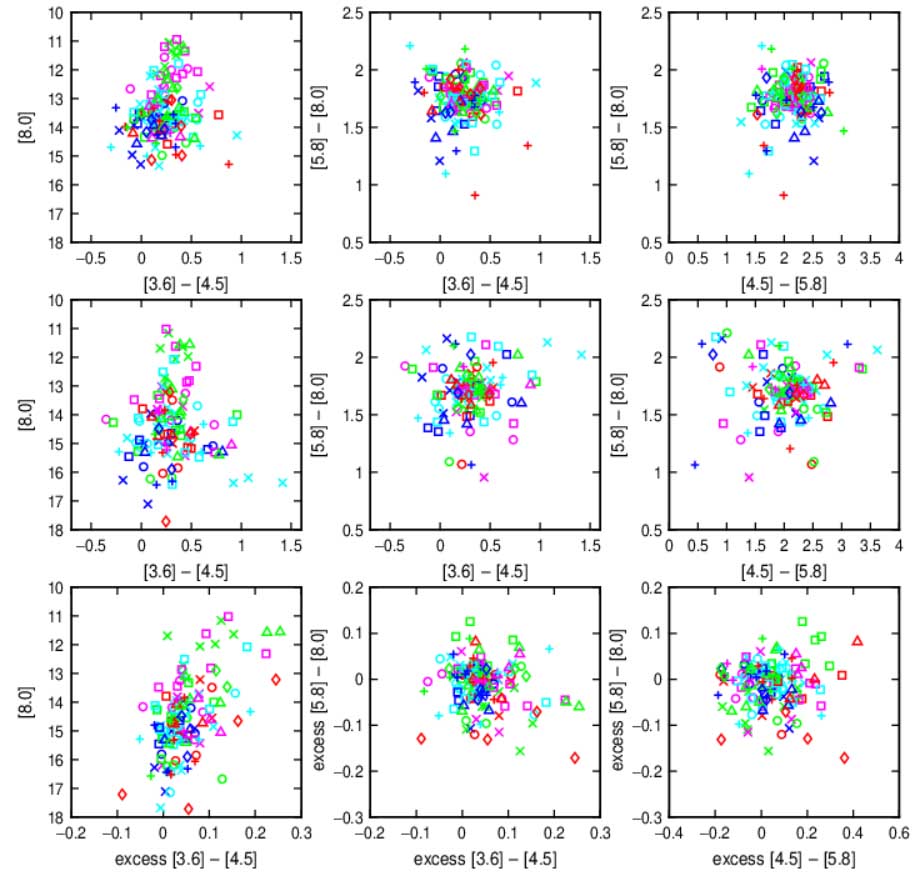}
\caption{Color-magnitude and color-color diagrams of the clumps in the filaments of
M100, using the same symbols as in figure
\ref{m100_spitzer_with_map_try7_nostars_cropped}
to indicate specific filaments. The units on both axes are magnitudes.
The top row calculates magnitudes and colors using
circular apertures of 2 pixels radius and background
subtraction using an annulus from 3 to 4 pixels away from the clump center.
The middle row determines the flux from each clump using a circular aperture of 1.5 pixel
radius and subtracts a background flux from the underlying filament, taken to be the average
of the fluxes in 1.5 pixel radius apertures on each side of the clump, midway to the next clump.
The magnitudes and colors of the clumps are calculated from these filament-subtracted fluxes.
The bottom row uses the same clump and average-filament fluxes as the middle row, but
the colors plotted are the excess colors of the clump compared to the average underlying filament.
}  \label{efremovbeads_cmd_minusmid_combine_colordiff}
\end{figure*}

\begin{figure*}
\epsscale{.7}
%\plotone{f8.eps}
\plotone{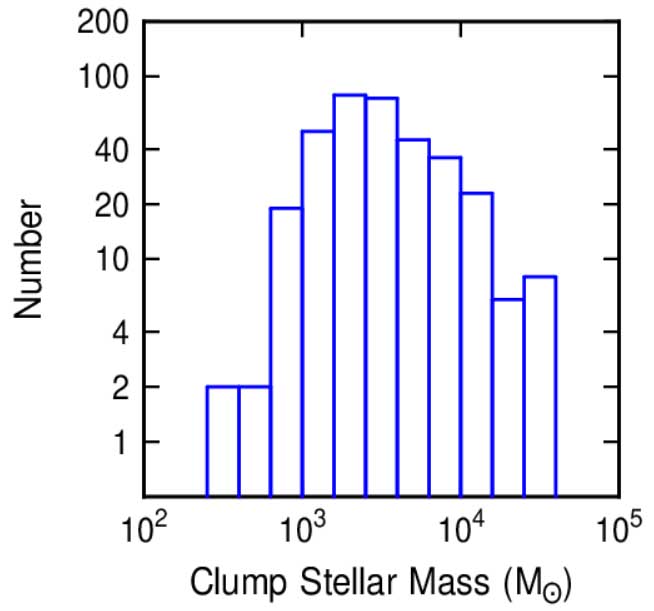}
\caption{Histogram showing the distribution of the equivalent stellar masses for heating of
all the measured clumps. The bolometric luminosity is assumed to equal 8 times the sum of the
IRAC luminosities, and the associated stellar mass assumes a bolometric magnitude per unit
mass equal to $-2.7944$, which is appropriate for ages less than 1 Myr. The masses would
all be larger by a factor of 5.1 if the ages are 10 Myr.
} \label{efremovbeads_mstaronly}
\end{figure*}

\begin{figure*}
\epsscale{.7}
%\plotone{f9.eps}
\plotone{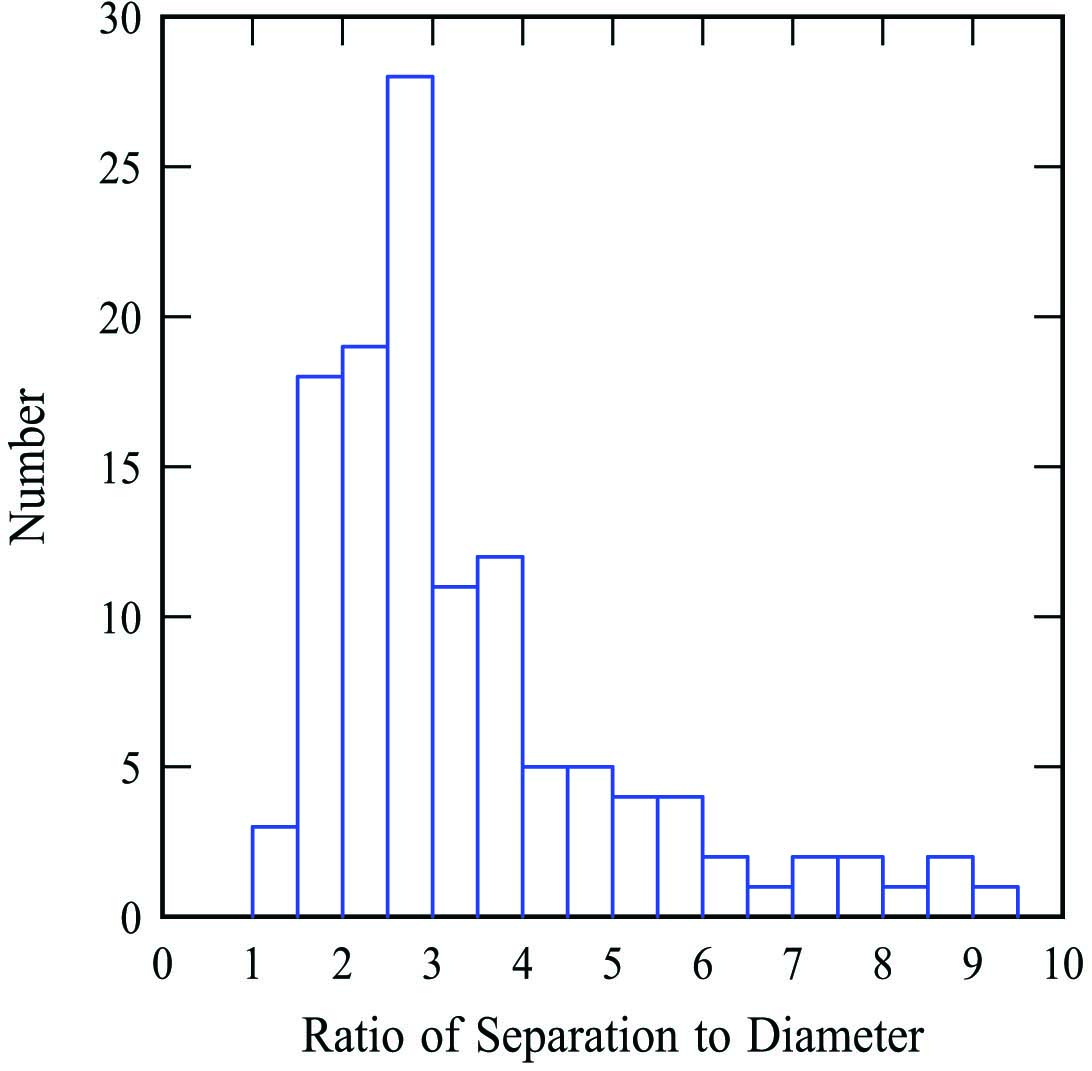}
\caption{Histogram showing the distribution of the ratio of the
separation between pairs of clumps to the average diameters of those clumps. The ratio
averages approximately 3, which is similar to the value expected from the fastest-growing
gravitational instability in a filament with a critical line density.
} \label{efremovbeads_ss}
\end{figure*}

Figure \ref{m100composite_clipped} shows good agreement with these expectations
for the self-gravitational collapse of filaments. The separations appear to be 3
to 5 times the filament diameters, as expected for near-critical line densities.
Figure \ref{efremovbeads_ss} shows this more quantitatively. It plots a histogram
of the ratio of the separation between adjacent clumps to the average clump
diameter. The diameters were determined from the number of pixels in rectangles
surrounding the clumps using the IRAF routine {\it imstat}. We consider that the
rectangles typically go to about 0.1 times the peak brightness and we assume the
diameters go to 0.5 times the peak brightness. In this case the diameters used for
Figure \ref{efremovbeads_ss} are about 1.1 times the square roots of the ratios of
the areas to $\pi$, measured in pixels like the separations. Figure
\ref{efremovbeads_ss} has a clear peak in the ratio of separation to diameter that
is centered at around 3, and a tail toward a value of $\sim6$ that could be from
adjacent clumps with a gap between them (see Fig. 4).  This is the ratio expected
if the clumps formed by gravitational instabilities.

The average gas densities inside the filaments are low because their apparent
sizes are large. For example, if the gas mass is $\sim100$ times the average
stellar mass in Figure \ref{efremovbeads_mstaronly}, which would be
$\sim3\times10^5\;M_\odot$ corresponding to a low star-formation efficiency, and
their diameters are typically $\sim130$ pc as measured from the surrounding box
sizes discussed above, then the average gas density is $\sim2.1$ atoms cm$^{-3}$.
Taking this density to be $\rho_c$ in the above equation, the corresponding
collapse time is $t_{\rm coll}=2.9(4\pi G\rho_{\rm c})^{-1/2}\sim46$ Myr.
This density is too low and the resulting time is too long to have a simultaneous
collapse of many clumps along each filament. The filaments probably last
only 10 Myr up to a few times 10 Myr, considering the distortions and shear of
spiral arms. Thus the average gas density inside the filaments is probably
much higher than what we see at the resolution of IRAC.

Dense filaments like the infrared dark clouds in the Milky Way \citep[IRDC;
e.g.][]{rathborne07,peretto09} could be at the cores of our IRAC filaments.
IRDC are opaque clouds observed against the bright background of diffuse IR
emission from the Galactic plane. There is a 160 to 430 pc long IRDC in the
Scutum-Centaurus Arm \citep{goodman14} that is a good candidate for a galactic
scale filament like the longer ones observed here. \cite{jackson10} also suggested
that a self-gravitational instability made regular condensations in the Milky Way
filament named ``Nessie,'' although that is on a much smaller scale with a 4.5 pc
clump separation and a total length of 80 pc.

The kiloparsec lengths of many IRAC filaments imply that dynamical processes
sweep up interstellar gas on this scale. For the main spiral arms, this process is
presumably the usual density wave shock, which can have a length of several kpc
for most of the regions inside corotation and possibly outside corotation too. The
main spiral arms in M100 are of this type, as evident from the similar positions
of broad stellar arms seen at $2\mu$m wavelength by the 2MASS survey
\citep{skrutskie06}
\footnote{https://www.ipac.caltech.edu/2mass/gallery/m100atlas.jpg}. Other
filaments could be from shear inside the stellar arms, making spurs (Section I).
Figures \ref{m100_ir_and_optical_rawforpub} to \ref{m100composite_clipped} show
more remote filaments too, with no apparent connection to the stellar arms seen in
the 2MASS image. These remote filaments suggest there are large-scale gas motions
independent of the main stellar spirals, possible from local instabilities or
stellar feedback.

If the evolution timescale for the filaments is $\sim10$ Myr, based on the
expectation of shear rates and spiral wave motions, and the total mass of the
measured clumps in the filaments is $4.5\times10^6\;M_\odot$ for this age, from
above, then the ratio of the mass to the age is $0.45\;M_\odot$ yr$^{-1}$, which
is lower than the total star formation rate in M100, $\sim2.6\;M_\odot$ yr$^{-1}$
\citep{kennicutt11}. For a 1 Myr timecale, the rate would be $0.89\;M_\odot$
yr$^{-1}$, which is still small. These small rates suggest that filament clumps
are not the only drivers of star formation in M100. This conclusion is consistent
with the other evidence given in the introduction that spiral waves affect the
total star formation rate by only small amounts. It is also consistent with the
appearance of many other star forming regions outside of the filaments. Still, the
appearance of highly regular clump structures in numerous long filaments suggest
that spiral arms trigger gas collapse and at least some cloud formation by
gravitational instabilities.

\section{Conclusions}

Dust filaments in M100 revealed by Spitzer IRAC images tend to have a regular
spacing of similar-mass clumps along their lengths, suggestive of a formation
process involving gravitational instabilities in gas that was accumulated by
the relative motion of spiral density waves and the associated large-scale flows.
The clump separation is typically 410 pc and the ratio of the separation to
the clump diameter ranges from 2 to 4. Many filaments extend for several kpc.
These regions appear to be galactic-scale analogs of local star-forming filaments
and filamentary IRDCs, and they probably form and evolve in a similar way, making
stars at the gravitating condensations. IRAC colors reflect their likely emission
from PAHs and hot dust, as modelled for galactic star-forming regions. The
effective stellar masses of the selected condensations average
$3\times10^3\;M_\odot$ for an age of 1 Myr, with a total effective mass in
the range of $0.9-4.5\times10^6\;M_\odot$ for the measured clumps in the filaments
if we assume ages of 1 Myr and 10 Myr, respectively. The ratio of these filament
clump masses to the assumed ages fall short of the star formation rate in M100 by
factors of 3 to 6, which is consistent with the relatively small influence that
spiral arms generally have on total star formation rates. The importance of the
observation lies in the identification of one process by which spiral waves
interact dynamically with the interstellar medium to form new clouds.

{\it Acknowledgments} This research has made use of the NASA/IPAC Extragalactic
Database (NED) which is operated by the Jet Propulsion Laboratory, California
Institute of Technology, under contract with the National Aeronautics and Space
Administration. This publication makes use of data products from the Two Micron
All Sky Survey, which is a joint project of the University of Massachusetts and
the Infrared Processing and Analysis Center/California Institute of Technology,
funded by the National Aeronautics and Space Administration and the National
Science Foundation. We are grateful to an anonymous referee for suggestions and to
Eric Feigelson for comments on our statistical procedures.

\newpage
%\begin{longtable}{llcccc}
\begin{deluxetable}{llcccc}
\tabletypesize{\scriptsize}\tablecolumns{6} \tablewidth{0pt}
\tablecaption{Clumps in Filaments with Background Annulus Subtracted}
\tablehead{
\colhead{RA}&
\colhead{DEC}&
\colhead{[3.6]}&
\colhead{[4.5]} &
\colhead{[5.8]} &
\colhead{[8.0]}\\
\colhead{}&
\colhead{}&
\colhead{mag}&
\colhead{mag} &
\colhead{mag} &
\colhead{mag}
}
\startdata
Filament  1, & blue {\it o} & & & & \\
12:22:46.4110&15:48:37.915&$17.86\pm 0.89$&$17.71\pm 1.03$&$15.48\pm 0.47$&$13.52\pm 0.26$\\
12:22:46.4110&15:48:34.165&$17.41\pm 0.72$&$17.18\pm 0.81$&$14.93\pm 0.36$&$13.16\pm 0.22$\\
12:22:46.3851&15:48:28.540&$17.58\pm 0.79$&$17.45\pm 0.92$&$15.35\pm 0.45$&$13.66\pm 0.31$\\
12:22:46.4371&15:48:25.915&$16.84\pm 0.57$&$16.43\pm 0.58$&$14.81\pm 0.36$&$13.08\pm 0.25$\\
Filament  2, & green triangle & & & & \\
12:22:47.9173&15:49:52.543&$18.57\pm 1.23$&$18.35\pm 1.38$&$15.60\pm 0.49$&$13.97\pm 0.34$\\
12:22:47.6055&15:49:48.418&$16.57\pm 0.49$&$16.31\pm 0.54$&$14.16\pm 0.26$&$12.34\pm 0.16$\\
12:22:47.4237&15:49:43.542&$15.33\pm 0.28$&$14.90\pm 0.28$&$13.03\pm 0.16$&$11.27\pm 0.11$\\
12:22:47.0600&15:49:31.542&$15.47\pm 0.30$&$15.11\pm 0.31$&$13.10\pm 0.16$&$11.38\pm 0.11$\\
12:22:46.9042&15:49:24.791&$16.93\pm 0.59$&$16.61\pm 0.63$&$14.34\pm 0.29$&$12.64\pm 0.20$\\
12:22:46.7483&15:49:22.166&$17.29\pm 0.69$&$17.13\pm 0.80$&$14.67\pm 0.33$&$13.00\pm 0.22$\\
12:22:46.7743&15:49:19.916&$17.44\pm 0.75$&$17.06\pm 0.77$&$15.18\pm 0.43$&$13.33\pm 0.25$\\
12:22:46.8003&15:49:16.166&$17.51\pm 0.76$&$17.25\pm 0.83$&$15.44\pm 0.46$&$13.62\pm 0.29$\\
12:22:46.3846&15:49:12.040&$18.27\pm 1.07$&$17.74\pm 1.05$&$16.21\pm 0.64$&$14.39\pm 0.40$\\
Filament  3, & red {\it o} & & & & \\
12:22:47.4508&15:47:51.792&$18.52\pm 1.20$&$18.38\pm 1.40$&$15.98\pm 0.58$&$14.14\pm 0.34$\\
12:22:47.6327&15:47:48.418&$18.06\pm 0.98$&$17.86\pm 1.11$&$16.06\pm 0.61$&$14.26\pm 0.41$\\
12:22:47.7367&15:47:43.543&$17.22\pm 0.66$&$16.95\pm 0.73$&$14.80\pm 0.34$&$13.03\pm 0.21$\\
Filament  4, & cyan {\it x} & & & & \\
12:22:49.5802&15:50:13.546&$17.42\pm 0.73$&$17.43\pm 0.92$&$14.94\pm 0.37$&$13.19\pm 0.23$\\
12:22:49.2165&15:50: 6.046&$17.81\pm 0.87$&$17.44\pm 0.91$&$15.40\pm 0.48$&$13.60\pm 0.35$\\
12:22:48.9826&15:50: 2.670&$18.37\pm 1.14$&$18.11\pm 1.25$&$15.92\pm 0.60$&$14.19\pm 0.41$\\
12:22:48.8787&15:49:58.920&$16.56\pm 0.49$&$16.47\pm 0.58$&$14.03\pm 0.24$&$12.21\pm 0.15$\\
12:22:48.7748&15:49:54.420&$17.25\pm 0.67$&$17.01\pm 0.75$&$14.98\pm 0.38$&$13.21\pm 0.24$\\
12:22:48.4630&15:49:52.169&$19.17\pm 1.65$&$18.21\pm 1.31$&$16.16\pm 0.69$&$14.27\pm 0.44$\\
12:22:48.2811&15:49:47.294&$16.76\pm 0.54$&$16.58\pm 0.62$&$14.41\pm 0.29$&$12.68\pm 0.18$\\
12:22:48.1252&15:49:45.419&$17.05\pm 0.61$&$16.89\pm 0.71$&$14.64\pm 0.34$&$12.93\pm 0.24$\\
12:22:47.8914&15:49:42.043&$17.57\pm 0.78$&$17.33\pm 0.87$&$14.79\pm 0.34$&$13.26\pm 0.26$\\
12:22:47.6057&15:49:30.793&$18.19\pm 1.03$&$17.74\pm 1.04$&$15.38\pm 0.44$&$13.67\pm 0.29$\\
12:22:47.1121&15:49:19.917&$17.89\pm 0.90$&$17.74\pm 1.04$&$15.53\pm 0.49$&$14.00\pm 0.39$\\
12:22:47.1641&15:49:17.667&$18.19\pm 1.04$&$18.07\pm 1.22$&$15.71\pm 0.56$&$13.94\pm 0.38$\\
12:22:47.1641&15:49:13.167&$17.26\pm 0.68$&$17.16\pm 0.80$&$15.06\pm 0.40$&$13.24\pm 0.25$\\
12:22:46.5925&15:49: 8.291&$18.53\pm 1.25$&$18.39\pm 1.45$&$16.49\pm 0.78$&$14.69\pm 0.53$\\
12:22:46.4887&15:49: 3.040&$18.05\pm 0.98$&$17.92\pm 1.13$&$15.69\pm 0.55$&$14.03\pm 0.38$\\
12:22:46.3588&15:49: 1.165&$18.31\pm 1.10$&$18.13\pm 1.25$&$16.88\pm 0.99$&$15.33\pm 0.79$\\
12:22:46.2289&15:48:58.165&$19.87\pm 2.33$&$21.50\pm 6.36$&$17.29\pm 1.29$&$15.91\pm 1.38$\\
Filament  5, & magenta triangle & & & & \\
12:22:50.2044&15:48:48.797&$17.42\pm 0.72$&$17.33\pm 0.86$&$15.13\pm 0.39$&$13.27\pm 0.23$\\
12:22:50.0226&15:48:37.922&$18.38\pm 1.12$&$18.27\pm 1.34$&$16.08\pm 0.61$&$14.38\pm 0.38$\\
12:22:50.3345&15:48:25.922&$18.99\pm 1.49$&$18.59\pm 1.55$&$16.17\pm 0.65$&$14.33\pm 0.37$\\
Filament  6, & green $+$ & & & & \\
12:22:51.1132&15:51: 0.423&$18.01\pm 0.95$&$18.15\pm 1.26$&$16.01\pm 0.65$&$14.00\pm 0.38$\\
12:22:50.9572&15:51: 3.798&$17.84\pm 0.89$&$17.60\pm 0.98$&$15.82\pm 0.56$&$13.64\pm 0.27$\\
12:22:50.6713&15:51: 6.048&$19.19\pm 1.68$&$19.05\pm 1.94$&$16.01\pm 0.59$&$14.54\pm 0.42$\\
Filament  7, & magenta {\it x}  & & & & \\
12:22:52.1528&15:50:24.049&$17.91\pm 0.91$&$17.33\pm 0.87$&$15.32\pm 0.43$&$13.50\pm 0.26$\\
12:22:51.6591&15:50:24.424&$17.10\pm 0.63$&$16.92\pm 0.72$&$14.56\pm 0.31$&$12.92\pm 0.21$\\
12:22:51.2433&15:50:24.798&$17.54\pm 0.77$&$16.86\pm 0.70$&$14.53\pm 0.31$&$12.58\pm 0.17$\\
12:22:50.8275&15:50:25.923&$18.39\pm 1.17$&$18.16\pm 1.27$&$15.69\pm 0.52$&$13.62\pm 0.32$\\
12:22:50.4897&15:50:24.048&$16.72\pm 0.53$&$16.51\pm 0.59$&$14.45\pm 0.29$&$12.67\pm 0.18$\\
Filament  8, & magenta $+$ & & & & \\
12:22:51.6849&15:51: 9.424&$17.20\pm 0.65$&$17.03\pm 0.75$&$14.89\pm 0.35$&$13.25\pm 0.23$\\
12:22:51.4250&15:51:14.299&$18.10\pm 0.99$&$17.81\pm 1.08$&$16.20\pm 0.67$&$14.19\pm 0.37$\\
12:22:51.1131&15:51:14.298&$18.31\pm 1.10$&$18.07\pm 1.22$&$15.91\pm 0.58$&$13.97\pm 0.32$\\
Filament  9, & magenta square & & & & \\
12:22:54.6476&15:49:33.050&$14.94\pm 0.27$&$14.71\pm 0.28$&$12.97\pm 0.18$&$11.20\pm 0.15$\\
12:22:54.0759&15:49:38.675&$16.51\pm 0.48$&$16.16\pm 0.51$&$14.02\pm 0.24$&$12.28\pm 0.15$\\
12:22:53.2444&15:49:39.800&$16.87\pm 0.56$&$16.77\pm 0.67$&$14.66\pm 0.33$&$12.87\pm 0.20$\\
12:22:52.5948&15:49:38.675&$16.16\pm 0.41$&$15.60\pm 0.39$&$13.75\pm 0.23$&$12.06\pm 0.17$\\
12:22:51.8672&15:49:35.299&$16.62\pm 0.51$&$16.21\pm 0.52$&$13.99\pm 0.26$&$12.14\pm 0.19$\\
12:22:51.6593&15:49:33.424&$15.56\pm 0.32$&$15.12\pm 0.32$&$13.21\pm 0.20$&$11.35\pm 0.14$\\
12:22:51.4255&15:49:30.424&$15.01\pm 0.25$&$14.66\pm 0.26$&$12.63\pm 0.14$&$10.95\pm 0.10$\\
12:22:51.1916&15:49:22.548&$16.31\pm 0.44$&$16.12\pm 0.50$&$14.03\pm 0.25$&$12.20\pm 0.15$\\
Filament 10, & cyan square & & & & \\
12:22:50.8020&15:48:57.798&$16.11\pm 0.40$&$15.98\pm 0.47$&$13.78\pm 0.22$&$12.02\pm 0.15$\\
12:22:51.5556&15:48:49.549&$16.88\pm 0.57$&$16.64\pm 0.63$&$14.56\pm 0.31$&$12.85\pm 0.20$\\
12:22:51.8674&15:48:44.299&$15.86\pm 0.35$&$15.49\pm 0.37$&$13.48\pm 0.19$&$11.78\pm 0.12$\\
12:22:52.5430&15:48:44.300&$18.11\pm 0.99$&$17.79\pm 1.07$&$15.49\pm 0.47$&$13.95\pm 0.32$\\
12:22:52.7768&15:48:39.425&$18.28\pm 1.09$&$18.02\pm 1.19$&$15.89\pm 0.57$&$13.84\pm 0.31$\\
12:22:52.9327&15:48:37.550&$18.32\pm 1.11$&$17.81\pm 1.09$&$16.12\pm 0.65$&$14.35\pm 0.46$\\
12:22:53.2186&15:48:32.300&$17.08\pm 0.62$&$17.00\pm 0.74$&$14.88\pm 0.36$&$13.19\pm 0.24$\\
12:22:53.1926&15:48:29.675&$17.81\pm 0.87$&$17.64\pm 1.01$&$15.02\pm 0.39$&$13.27\pm 0.26$\\
12:22:54.1280&15:48:28.550&$17.78\pm 0.88$&$17.85\pm 1.12$&$15.48\pm 0.47$&$13.48\pm 0.27$\\
12:22:54.1539&15:48:24.050&$17.35\pm 0.71$&$16.76\pm 0.66$&$14.75\pm 0.36$&$12.86\pm 0.21$\\
12:22:54.2319&15:48:21.050&$17.12\pm 0.63$&$16.78\pm 0.67$&$15.05\pm 0.44$&$13.76\pm 0.42$\\
Filament 11, & red $+$ & & & & \\
12:22:54.0499&15:50:16.175&$19.15\pm 1.66$&$18.27\pm 1.34$&$16.62\pm 0.88$&$15.28\pm 0.72$\\
12:22:53.6601&15:50:18.800&$18.41\pm 1.16$&$18.57\pm 1.56$&$15.78\pm 0.54$&$13.98\pm 0.33$\\
12:22:53.3482&15:50:22.175&$18.20\pm 1.04$&$17.85\pm 1.10$&$15.86\pm 0.56$&$14.95\pm 0.52$\\
12:22:52.9584&15:50:24.050&$17.57\pm 0.78$&$17.44\pm 0.91$&$15.16\pm 0.40$&$13.32\pm 0.25$\\
12:22:52.7246&15:50:21.800&$18.20\pm 1.04$&$18.01\pm 1.18$&$15.79\pm 0.54$&$13.78\pm 0.30$\\
Filament 12, & green square & & & & \\
12:22:55.4791&15:49:54.050&$17.20\pm 0.65$&$17.08\pm 0.77$&$15.08\pm 0.38$&$13.28\pm 0.23$\\
12:22:54.6736&15:49:58.925&$17.72\pm 0.84$&$17.42\pm 0.90$&$15.62\pm 0.49$&$13.64\pm 0.27$\\
12:22:54.3877&15:50: 1.550&$18.04\pm 0.98$&$17.62\pm 0.99$&$15.35\pm 0.44$&$13.46\pm 0.25$\\
12:22:54.1019&15:50: 4.550&$17.16\pm 0.66$&$16.99\pm 0.75$&$15.00\pm 0.39$&$13.09\pm 0.24$\\
12:22:53.9200&15:50: 2.300&$17.70\pm 0.83$&$17.64\pm 1.01$&$15.40\pm 0.52$&$13.47\pm 0.34$\\
12:22:53.6341&15:50: 0.425&$17.73\pm 0.85$&$17.41\pm 0.91$&$15.38\pm 0.49$&$13.62\pm 0.35$\\
12:22:53.1144&15:50: 3.800&$17.47\pm 0.74$&$17.44\pm 0.91$&$16.02\pm 0.61$&$14.35\pm 0.41$\\
12:22:52.7506&15:50: 5.300&$18.57\pm 1.32$&$18.32\pm 1.39$&$15.69\pm 0.51$&$13.78\pm 0.29$\\
12:22:52.4388&15:50: 2.300&$18.25\pm 1.07$&$18.33\pm 1.38$&$15.69\pm 0.50$&$13.70\pm 0.28$\\
12:22:51.7371&15:50: 6.049&$17.59\pm 0.79$&$17.04\pm 0.76$&$15.39\pm 0.45$&$13.67\pm 0.29$\\
Filament 13, & green {\it o} & & & & \\
12:22:54.6735&15:51:23.300&$18.62\pm 1.26$&$18.29\pm 1.34$&$15.78\pm 0.53$&$14.01\pm 0.32$\\
12:22:54.3617&15:51:25.925&$18.87\pm 1.42$&$18.43\pm 1.43$&$16.11\pm 0.62$&$14.54\pm 0.42$\\
12:22:54.0758&15:51:27.050&$19.93\pm 2.30$&$19.72\pm 2.60$&$17.03\pm 0.95$&$14.98\pm 0.51$\\
12:22:53.6340&15:51:30.425&$18.35\pm 1.12$&$18.17\pm 1.28$&$15.75\pm 0.53$&$14.09\pm 0.34$\\
Filament 14, & green diamond & & & & \\
12:22:55.9730&15:51: 2.675&$16.48\pm 0.47$&$16.11\pm 0.49$&$14.24\pm 0.27$&$12.62\pm 0.19$\\
12:22:55.8170&15:51: 7.550&$17.60\pm 0.79$&$17.60\pm 0.98$&$15.26\pm 0.42$&$13.50\pm 0.26$\\
12:22:55.5572&15:51:13.550&$16.32\pm 0.44$&$16.07\pm 0.48$&$14.07\pm 0.24$&$12.14\pm 0.14$\\
Filament 15, & green {\it x}  & & & & \\
12:22:57.7135&15:48: 6.424&$15.17\pm 0.26$&$14.89\pm 0.28$&$12.76\pm 0.15$&$11.05\pm 0.11$\\
12:22:57.1419&15:48: 6.800&$15.33\pm 0.29$&$14.94\pm 0.30$&$12.85\pm 0.16$&$11.14\pm 0.12$\\
12:22:56.6222&15:48: 7.925&$15.91\pm 0.37$&$15.75\pm 0.43$&$13.84\pm 0.29$&$12.14\pm 0.25$\\
12:22:56.2325&15:48:10.175&$15.51\pm 0.30$&$15.21\pm 0.33$&$13.06\pm 0.16$&$11.34\pm 0.12$\\
12:22:55.8688&15:48:11.675&$15.68\pm 0.33$&$15.33\pm 0.35$&$13.20\pm 0.18$&$11.49\pm 0.13$\\
12:22:55.3751&15:48:15.050&$16.03\pm 0.38$&$15.79\pm 0.43$&$13.65\pm 0.20$&$11.87\pm 0.13$\\
12:22:54.7515&15:48:16.925&$15.07\pm 0.25$&$14.80\pm 0.27$&$12.73\pm 0.14$&$11.06\pm 0.11$\\
Filament 16, & blue square & & & & \\
12:22:56.7783&15:49: 4.550&$18.19\pm 1.03$&$18.21\pm 1.29$&$15.51\pm 0.47$&$13.57\pm 0.26$\\
12:22:57.3239&15:49: 4.550&$17.06\pm 0.61$&$17.10\pm 0.78$&$15.52\pm 0.50$&$13.87\pm 0.39$\\
12:22:57.5578&15:49: 6.049&$17.29\pm 0.69$&$17.05\pm 0.76$&$15.21\pm 0.41$&$13.68\pm 0.31$\\
12:22:57.8956&15:49: 9.049&$17.80\pm 0.87$&$17.82\pm 1.10$&$15.37\pm 0.45$&$13.52\pm 0.29$\\
Filament 17, & red square & & & & \\
12:22:58.0775&15:49:19.549&$19.10\pm 1.62$&$18.84\pm 1.74$&$16.61\pm 0.77$&$14.59\pm 0.42$\\
12:22:57.8437&15:49:25.924&$18.09\pm 1.00$&$17.77\pm 1.06$&$15.47\pm 0.46$&$13.88\pm 0.32$\\
12:22:57.6618&15:49:25.174&$18.80\pm 1.40$&$18.02\pm 1.20$&$15.38\pm 0.44$&$13.56\pm 0.27$\\
12:22:57.4020&15:49:31.175&$17.74\pm 0.84$&$17.70\pm 1.03$&$15.33\pm 0.44$&$13.48\pm 0.26$\\
Filament 18, & red diamond & & & & \\
12:22:58.3898&15:50:57.799&$17.38\pm 0.71$&$17.08\pm 0.77$&$14.83\pm 0.34$&$13.05\pm 0.21$\\
12:22:57.8441&15:51: 4.924&$18.41\pm 1.14$&$18.01\pm 1.18$&$15.79\pm 0.53$&$13.95\pm 0.31$\\
12:22:57.5323&15:51: 7.924&$18.52\pm 1.20$&$18.12\pm 1.24$&$16.58\pm 0.78$&$14.97\pm 0.53$\\
12:22:57.2724&15:51: 9.800&$19.34\pm 1.75$&$19.23\pm 2.07$&$17.05\pm 0.96$&$15.13\pm 0.54$\\
Filament 19, & cyan {\it o} & & & & \\
12:22:59.1169&15:49:19.923&$18.08\pm 0.99$&$18.02\pm 1.19$&$15.69\pm 0.52$&$13.75\pm 0.31$\\
12:22:59.0390&15:49:28.173&$17.89\pm 0.91$&$17.72\pm 1.04$&$15.33\pm 0.43$&$13.55\pm 0.26$\\
12:22:58.7012&15:49:30.423&$16.89\pm 0.57$&$16.32\pm 0.55$&$14.65\pm 0.35$&$13.12\pm 0.32$\\
12:22:58.3634&15:49:35.299&$17.18\pm 0.65$&$17.13\pm 0.79$&$15.21\pm 0.41$&$13.62\pm 0.30$\\
12:22:58.1036&15:49:40.174&$17.50\pm 0.75$&$17.10\pm 0.78$&$15.18\pm 0.40$&$13.42\pm 0.24$\\
12:22:57.8698&15:49:42.424&$18.81\pm 1.38$&$21.21\pm 6.01$&$16.78\pm 0.85$&$14.68\pm 0.45$\\
12:22:57.5320&15:49:49.174&$18.20\pm 1.06$&$17.80\pm 1.08$&$15.76\pm 0.54$&$13.80\pm 0.31$\\
12:22:57.3761&15:49:51.050&$18.37\pm 1.16$&$17.81\pm 1.10$&$15.78\pm 0.53$&$13.74\pm 0.28$\\
Filament 20, & blue $+$ & & & & \\
12:22:59.1427&15:48:39.798&$18.39\pm 1.15$&$18.04\pm 1.21$&$15.54\pm 0.47$&$13.86\pm 0.32$\\
12:22:58.9348&15:48:39.798&$18.33\pm 1.13$&$17.98\pm 1.19$&$16.47\pm 0.87$&$14.69\pm 0.68$\\
12:22:58.5970&15:48:40.923&$17.40\pm 0.73$&$17.24\pm 0.84$&$15.54\pm 0.50$&$14.25\pm 0.49$\\
12:22:58.3372&15:48:41.299&$17.73\pm 0.87$&$17.99\pm 1.24$&$15.21\pm 0.42$&$13.32\pm 0.24$\\
12:22:58.1034&15:48:45.049&$17.65\pm 0.81$&$17.33\pm 0.87$&$15.29\pm 0.44$&$13.56\pm 0.31$\\
Filament 21, & magenta {\it o} & & & & \\
12:22:59.5585&15:48:54.797&$15.82\pm 0.35$&$15.59\pm 0.39$&$13.28\pm 0.17$&$11.56\pm 0.11$\\
12:22:59.2987&15:48:54.798&$16.66\pm 0.57$&$16.43\pm 0.62$&$14.36\pm 0.40$&$12.60\pm 0.35$\\
12:22:59.0128&15:48:54.048&$16.91\pm 0.68$&$17.02\pm 0.90$&$14.60\pm 0.52$&$12.66\pm 0.38$\\
12:22:58.7790&15:48:55.923&$16.38\pm 0.47$&$16.08\pm 0.50$&$13.62\pm 0.22$&$12.00\pm 0.19$\\
12:22:58.5971&15:48:56.673&$15.91\pm 0.37$&$15.45\pm 0.37$&$13.62\pm 0.22$&$11.95\pm 0.17$\\
12:22:58.3633&15:48:57.424&$16.76\pm 0.54$&$16.50\pm 0.59$&$14.05\pm 0.24$&$12.30\pm 0.17$\\
Filament 22, & blue diamond & & & & \\
12:23: 0.9365&15:50:45.045&$17.76\pm 0.85$&$17.66\pm 1.00$&$15.36\pm 0.44$&$13.73\pm 0.28$\\
12:23: 0.8066&15:50:49.545&$17.93\pm 0.92$&$17.70\pm 1.02$&$16.00\pm 0.60$&$14.07\pm 0.33$\\
12:23: 0.6507&15:50:52.546&$18.48\pm 1.18$&$18.43\pm 1.44$&$15.79\pm 0.53$&$14.16\pm 0.35$\\
Filament 23, & red {\it x}  & & & & \\
12:23: 1.3263&15:50:40.169&$16.89\pm 0.57$&$16.70\pm 0.65$&$14.38\pm 0.28$&$12.69\pm 0.18$\\
12:23: 1.3783&15:50:46.544&$17.50\pm 0.75$&$17.28\pm 0.84$&$15.11\pm 0.41$&$13.38\pm 0.27$\\
12:23: 1.5083&15:50:49.169&$17.05\pm 0.61$&$16.88\pm 0.71$&$14.78\pm 0.34$&$12.92\pm 0.19$\\
Filament 24, & red triangle & & & & \\
12:23: 1.4819&15:50: 0.419&$18.14\pm 1.01$&$18.23\pm 1.31$&$15.84\pm 0.55$&$14.20\pm 0.35$\\
12:23: 1.3520&15:50: 8.294&$18.08\pm 0.98$&$17.89\pm 1.11$&$15.80\pm 0.53$&$14.08\pm 0.34$\\
12:23: 1.4821&15:50:21.044&$18.47\pm 1.17$&$18.27\pm 1.33$&$15.96\pm 0.58$&$14.12\pm 0.36$\\
Filament 25, & blue {\it x}  & & & & \\
12:23: 2.4698&15:50:42.792&$18.19\pm 1.03$&$18.41\pm 1.42$&$15.92\pm 0.58$&$14.10\pm 0.35$\\
12:23: 2.5998&15:50:52.917&$17.49\pm 0.75$&$17.37\pm 0.88$&$15.42\pm 0.45$&$13.64\pm 0.28$\\
12:23: 2.5999&15:50:58.542&$19.00\pm 1.50$&$19.01\pm 1.87$&$16.49\pm 0.76$&$15.28\pm 0.62$\\
12:23: 2.4960&15:51: 3.792&$19.05\pm 1.53$&$19.14\pm 1.99$&$16.53\pm 0.76$&$14.95\pm 0.51$\\
12:23: 2.4181&15:51: 8.667&$17.86\pm 0.89$&$17.94\pm 1.14$&$15.91\pm 0.56$&$13.92\pm 0.31$\\
Filament 26, & blue triangle & & & & \\
12:23: 2.6245&15:48:48.417&$17.88\pm 0.89$&$17.75\pm 1.05$&$15.40\pm 0.44$&$13.93\pm 0.32$\\
12:23: 2.8324&15:48:50.666&$18.82\pm 1.40$&$18.61\pm 1.57$&$16.01\pm 0.59$&$14.29\pm 0.40$\\
12:23: 3.2222&15:48:54.790&$18.11\pm 1.00$&$18.16\pm 1.27$&$15.98\pm 0.59$&$14.57\pm 0.41$\\
Filament 27, & cyan $+$ & & & & \\
12:23: 3.7676&15:48:36.039&$18.20\pm 1.04$&$18.51\pm 1.50$&$16.90\pm 0.90$&$14.69\pm 0.48$\\
12:23: 4.0795&15:48:45.413&$17.54\pm 0.77$&$17.49\pm 0.93$&$16.10\pm 0.62$&$15.01\pm 0.55$\\
12:23: 4.3654&15:48:51.038&$17.92\pm 0.91$&$17.48\pm 0.93$&$15.69\pm 0.51$&$13.88\pm 0.31$\\
12:23: 4.5734&15:48:58.162&$19.06\pm 1.54$&$18.47\pm 1.46$&$16.28\pm 0.66$&$14.64\pm 0.43$\\
\enddata
%\tablenotetext{1}{In order of average RA}
%\tablenotetext{2}{in magnitudes}
\label{tab}
\end{deluxetable}
%\end{longtable}

\newpage
%\begin{longtable}{llcccc}
\begin{deluxetable}{llcccc}
\tabletypesize{\scriptsize}\tablecolumns{6} \tablewidth{0pt}
\tablecaption{Clumps in Filaments with Interclump Subtracted}
\tablehead{
\colhead{RA}&
\colhead{DEC}&
\colhead{[3.6]}&
\colhead{[4.5]} &
\colhead{[5.8]} &
\colhead{[8.0]}\\
\colhead{}&
\colhead{}&
\colhead{mag}&
\colhead{mag} &
\colhead{mag} &
\colhead{mag}
}
\startdata
Filament  1, & blue {\it o} & & & & \\
12:22:46.4110&15:48:34.165&$18.64\pm 0.85$&$18.28\pm 1.01$&$15.95\pm 0.49$&$14.19\pm 0.30$\\
12:22:46.3851&15:48:28.540&$19.70\pm 0.86$&$19.68\pm 1.04$&$17.71\pm 0.53$&$15.80\pm 0.33$\\
12:22:46.4371&15:48:25.915&$18.97\pm 0.74$&$18.25\pm 0.83$&$16.69\pm 0.44$&$15.08\pm 0.28$\\
Filament  2, & green triangle & & & & \\
12:22:47.6055&15:49:48.418&$17.37\pm 0.62$&$17.06\pm 0.71$&$14.92\pm 0.34$&$13.10\pm 0.20$\\
12:22:47.4237&15:49:43.542&$15.69\pm 0.53$&$15.20\pm 0.60$&$13.24\pm 0.29$&$11.55\pm 0.18$\\
12:22:47.0600&15:49:31.542&$15.75\pm 0.57$&$15.36\pm 0.65$&$13.28\pm 0.31$&$11.57\pm 0.18$\\
12:22:46.9042&15:49:24.791&$17.89\pm 0.68$&$17.49\pm 0.77$&$15.59\pm 0.37$&$13.89\pm 0.22$\\
12:22:46.7483&15:49:22.166&$19.31\pm 0.75$&$18.94\pm 0.85$&$17.00\pm 0.41$&$15.46\pm 0.26$\\
12:22:46.7743&15:49:19.916&$19.57\pm 0.79$&$18.80\pm 0.89$&$17.41\pm 0.47$&$15.39\pm 0.29$\\
12:22:46.8003&15:49:16.166&$18.37\pm 0.83$&$18.17\pm 1.00$&$16.05\pm 0.52$&$14.14\pm 0.32$\\
12:22:46.3846&15:49:12.040&$18.21\pm 0.89$&$17.82\pm 1.07$&$15.84\pm 0.57$&$14.02\pm 0.36$\\
Filament  3, & red {\it o} & & & & \\
12:22:47.4508&15:47:51.792&$19.80\pm 1.18$&$19.58\pm 1.45$&$17.10\pm 0.71$&$16.03\pm 0.45$\\
12:22:47.6327&15:47:48.418&$19.01\pm 1.06$&$18.64\pm 1.26$&$17.76\pm 0.70$&$15.85\pm 0.41$\\
12:22:47.7367&15:47:43.543&$17.72\pm 0.89$&$17.41\pm 1.04$&$15.27\pm 0.50$&$13.49\pm 0.30$\\
Filament  4, & cyan {\it x} & & & & \\
12:22:49.5802&15:50:13.546&$18.19\pm 0.67$&$18.03\pm 0.82$&$15.68\pm 0.44$&$14.07\pm 0.27$\\
12:22:49.2165&15:50: 6.046&$18.63\pm 0.69$&$18.47\pm 0.85$&$16.20\pm 0.46$&$14.43\pm 0.28$\\
12:22:48.8787&15:49:58.920&$17.28\pm 0.58$&$17.06\pm 0.69$&$14.83\pm 0.34$&$13.02\pm 0.21$\\
12:22:48.7748&15:49:54.420&$18.32\pm 0.68$&$18.05\pm 0.81$&$16.70\pm 0.45$&$14.95\pm 0.28$\\
12:22:48.4630&15:49:52.169&$22.54\pm 0.90$&$21.13\pm 1.02$&$18.38\pm 0.55$&$16.36\pm 0.34$\\
12:22:48.2811&15:49:47.294&$17.81\pm 0.63$&$17.48\pm 0.74$&$15.61\pm 0.39$&$13.94\pm 0.25$\\
12:22:48.1252&15:49:45.419&$20.35\pm 0.73$&$20.49\pm 0.88$&$16.87\pm 0.41$&$14.81\pm 0.25$\\
12:22:47.8914&15:49:42.043&$18.47\pm 0.76$&$18.21\pm 0.90$&$15.65\pm 0.46$&$13.94\pm 0.29$\\
12:22:47.6057&15:49:30.793&$19.00\pm 0.87$&$18.60\pm 1.02$&$16.20\pm 0.54$&$14.44\pm 0.33$\\
12:22:47.1121&15:49:19.917&$19.65\pm 0.89$&$19.15\pm 1.04$&$16.77\pm 0.51$&$14.97\pm 0.32$\\
12:22:47.1641&15:49:17.667&$20.57\pm 0.94$&$19.65\pm 1.08$&$19.49\pm 0.59$&$16.37\pm 0.34$\\
12:22:47.1641&15:49:13.167&$18.05\pm 0.85$&$17.79\pm 1.03$&$15.52\pm 0.51$&$13.76\pm 0.31$\\
12:22:46.5925&15:49: 8.291&$18.86\pm 0.98$&$18.39\pm 1.18$&$16.76\pm 0.65$&$14.85\pm 0.39$\\
12:22:46.4887&15:49: 3.040&$19.53\pm 0.96$&$19.27\pm 1.19$&$17.16\pm 0.61$&$15.35\pm 0.37$\\
12:22:46.3588&15:49: 1.165&$21.17\pm 1.05$&$20.09\pm 1.22$&$18.32\pm 0.63$&$16.19\pm 0.38$\\
Filament  5, & magenta triangle & & & & \\
12:22:50.2044&15:48:48.797&$17.90\pm 0.86$&$17.71\pm 1.05$&$15.58\pm 0.56$&$13.87\pm 0.35$\\
12:22:50.0226&15:48:37.922&$19.15\pm 1.04$&$18.66\pm 1.26$&$16.32\pm 0.66$&$14.61\pm 0.42$\\
12:22:50.3345&15:48:25.922&$20.07\pm 1.21$&$19.17\pm 1.43$&$16.81\pm 0.76$&$15.05\pm 0.49$\\
Filament  6, & green $+$ & & & & \\
12:22:50.9572&15:51: 3.798&$18.81\pm 0.94$&$18.68\pm 1.13$&$16.77\pm 0.60$&$14.67\pm 0.36$\\
Filament  7, & magenta {\it x}  & & & & \\
12:22:52.1528&15:50:24.049&$18.21\pm 0.70$&$17.77\pm 0.83$&$16.38\pm 0.50$&$15.43\pm 0.32$\\
12:22:51.6591&15:50:24.424&$17.71\pm 0.65$&$17.55\pm 0.79$&$15.34\pm 0.42$&$13.81\pm 0.26$\\
12:22:51.2433&15:50:24.798&$17.99\pm 0.64$&$17.57\pm 0.76$&$15.48\pm 0.40$&$13.70\pm 0.24$\\
12:22:50.4897&15:50:24.048&$17.03\pm 0.56$&$16.82\pm 0.66$&$14.77\pm 0.34$&$13.02\pm 0.20$\\
Filament  8, & magenta $+$ & & & & \\
12:22:51.6849&15:51: 9.424&$17.66\pm 0.94$&$17.48\pm 1.10$&$15.63\pm 0.53$&$13.90\pm 0.33$\\
12:22:51.4250&15:51:14.299&$18.60\pm 1.05$&$18.37\pm 1.24$&$16.92\pm 0.63$&$15.00\pm 0.38$\\
12:22:51.1131&15:51:14.298&$18.99\pm 1.09$&$18.74\pm 1.31$&$16.96\pm 0.66$&$15.22\pm 0.40$\\
Filament  9, & magenta square & & & & \\
12:22:54.6476&15:49:33.050&$14.81\pm 0.32$&$14.56\pm 0.38$&$12.72\pm 0.23$&$11.02\pm 0.14$\\
12:22:54.0759&15:49:38.675&$17.18\pm 0.47$&$16.70\pm 0.55$&$14.51\pm 0.32$&$12.81\pm 0.20$\\
12:22:53.2444&15:49:39.800&$17.70\pm 0.55$&$17.52\pm 0.67$&$15.12\pm 0.37$&$13.44\pm 0.23$\\
12:22:52.5948&15:49:38.675&$16.52\pm 0.49$&$15.98\pm 0.56$&$13.99\pm 0.31$&$12.31\pm 0.19$\\
12:22:51.8672&15:49:35.299&$18.62\pm 0.53$&$18.69\pm 0.63$&$15.39\pm 0.31$&$13.48\pm 0.18$\\
12:22:51.6593&15:49:33.424&$17.02\pm 0.40$&$16.57\pm 0.44$&$14.98\pm 0.23$&$12.87\pm 0.14$\\
12:22:51.4255&15:49:30.424&$15.63\pm 0.36$&$15.29\pm 0.41$&$13.29\pm 0.20$&$11.61\pm 0.12$\\
12:22:51.1916&15:49:22.548&$18.40\pm 0.51$&$17.67\pm 0.60$&$16.73\pm 0.34$&$15.31\pm 0.21$\\
Filament 10, & cyan square & & & & \\
12:22:50.8020&15:48:57.798&$16.88\pm 0.55$&$16.66\pm 0.65$&$14.24\pm 0.32$&$12.51\pm 0.20$\\
12:22:51.5556&15:48:49.549&$17.46\pm 0.58$&$17.49\pm 0.71$&$15.55\pm 0.37$&$14.11\pm 0.23$\\
12:22:51.8674&15:48:44.299&$16.13\pm 0.51$&$15.80\pm 0.61$&$13.77\pm 0.32$&$12.08\pm 0.19$\\
12:22:52.5430&15:48:44.300&$19.46\pm 0.74$&$19.03\pm 0.90$&$16.76\pm 0.50$&$15.32\pm 0.32$\\
12:22:52.7768&15:48:39.425&$20.14\pm 0.76$&$19.56\pm 0.92$&$17.57\pm 0.53$&$15.48\pm 0.32$\\
12:22:53.2186&15:48:32.300&$18.18\pm 0.65$&$17.98\pm 0.78$&$15.95\pm 0.41$&$14.26\pm 0.26$\\
12:22:54.1280&15:48:28.550&$19.72\pm 0.74$&$19.41\pm 0.91$&$18.60\pm 0.54$&$16.42\pm 0.33$\\
12:22:54.1539&15:48:24.050&$18.84\pm 0.65$&$17.92\pm 0.75$&$16.09\pm 0.39$&$14.24\pm 0.24$\\
12:22:54.2319&15:48:21.050&$17.95\pm 0.59$&$17.79\pm 0.70$&$16.74\pm 0.38$&$15.04\pm 0.24$\\
Filament 11, & red $+$ & & & & \\
12:22:54.0499&15:50:16.175&$22.55\pm 0.90$&$19.37\pm 0.95$&$17.26\pm 0.56$&$16.06\pm 0.37$\\
12:22:53.6601&15:50:18.800&$20.01\pm 0.78$&$19.49\pm 0.94$&$16.63\pm 0.53$&$14.67\pm 0.32$\\
12:22:53.3482&15:50:22.175&$18.82\pm 0.72$&$18.60\pm 0.88$&$16.96\pm 0.53$&$15.32\pm 0.34$\\
12:22:52.9584&15:50:24.050&$18.06\pm 0.67$&$17.82\pm 0.81$&$15.60\pm 0.44$&$13.84\pm 0.27$\\
12:22:52.7246&15:50:21.800&$23.92\pm 0.90$&$20.77\pm 0.97$&$20.40\pm 0.59$&$16.51\pm 0.32$\\
Filament 12, & green square & & & & \\
12:22:55.4791&15:49:54.050&$17.29\pm 0.60$&$17.13\pm 0.74$&$15.04\pm 0.45$&$13.33\pm 0.29$\\
12:22:54.6736&15:49:58.925&$18.32\pm 0.65$&$18.05\pm 0.80$&$15.81\pm 0.48$&$14.04\pm 0.31$\\
12:22:54.3877&15:50: 1.550&$19.13\pm 0.68$&$18.74\pm 0.82$&$16.26\pm 0.48$&$14.52\pm 0.30$\\
12:22:54.1019&15:50: 4.550&$18.92\pm 0.65$&$18.58\pm 0.78$&$16.50\pm 0.44$&$14.54\pm 0.27$\\
12:22:53.6341&15:50: 0.425&$19.37\pm 0.69$&$18.97\pm 0.83$&$16.25\pm 0.48$&$14.73\pm 0.31$\\
12:22:53.1144&15:50: 3.800&$19.02\pm 0.68$&$19.06\pm 0.84$&$16.96\pm 0.53$&$15.29\pm 0.34$\\
12:22:52.4388&15:50: 2.300&$19.25\pm 0.72$&$19.53\pm 0.93$&$16.17\pm 0.53$&$14.28\pm 0.34$\\
12:22:51.7371&15:50: 6.049&$19.02\pm 0.70$&$18.07\pm 0.81$&$15.79\pm 0.50$&$14.00\pm 0.33$\\
Filament 13, & green {\it o} & & & & \\
12:22:54.3617&15:51:25.925&$19.93\pm 1.26$&$19.84\pm 1.54$&$17.32\pm 0.73$&$16.23\pm 0.47$\\
12:22:54.0758&15:51:27.050&$20.61\pm 1.38$&$20.31\pm 1.66$&$17.73\pm 0.80$&$16.18\pm 0.49$\\
12:22:53.6340&15:51:30.425&$18.60\pm 1.18$&$18.43\pm 1.43$&$16.09\pm 0.69$&$14.54\pm 0.43$\\
Filament 14, & green diamond & & & & \\
12:22:55.9730&15:51: 2.675&$17.35\pm 0.72$&$16.93\pm 0.83$&$15.19\pm 0.41$&$13.47\pm 0.25$\\
Filament 15, & green {\it x}  & & & & \\
12:22:57.7135&15:48: 6.424&$15.54\pm 0.39$&$15.35\pm 0.45$&$13.29\pm 0.21$&$11.69\pm 0.13$\\
12:22:57.1419&15:48: 6.800&$15.77\pm 0.42$&$15.35\pm 0.48$&$13.28\pm 0.22$&$11.63\pm 0.13$\\
12:22:56.6222&15:48: 7.925&$16.64\pm 0.49$&$16.38\pm 0.56$&$14.72\pm 0.27$&$13.18\pm 0.17$\\
12:22:56.2325&15:48:10.175&$16.18\pm 0.43$&$15.83\pm 0.48$&$13.76\pm 0.22$&$12.06\pm 0.14$\\
12:22:55.8688&15:48:11.675&$16.58\pm 0.44$&$16.11\pm 0.50$&$13.70\pm 0.23$&$11.98\pm 0.14$\\
12:22:55.3751&15:48:15.050&$16.83\pm 0.50$&$16.59\pm 0.59$&$14.38\pm 0.29$&$12.72\pm 0.18$\\
12:22:54.7515&15:48:16.925&$15.19\pm 0.44$&$14.92\pm 0.51$&$12.83\pm 0.24$&$11.15\pm 0.14$\\
Filament 16, & blue square & & & & \\
12:22:56.7783&15:49: 4.550&$18.66\pm 0.60$&$18.79\pm 0.75$&$16.84\pm 0.47$&$15.46\pm 0.30$\\
12:22:57.3239&15:49: 4.550&$17.79\pm 0.55$&$17.81\pm 0.68$&$16.23\pm 0.42$&$14.88\pm 0.27$\\
12:22:57.5578&15:49: 6.049&$18.09\pm 0.56$&$17.80\pm 0.67$&$16.14\pm 0.41$&$14.65\pm 0.26$\\
12:22:57.8956&15:49: 9.049&$19.53\pm 0.61$&$18.87\pm 0.74$&$17.23\pm 0.46$&$15.21\pm 0.28$\\
Filament 17, & red square & & & & \\
12:22:58.0775&15:49:19.549&$18.82\pm 0.67$&$18.36\pm 0.82$&$16.82\pm 0.55$&$15.15\pm 0.34$\\
12:22:57.8437&15:49:25.924&$19.18\pm 0.67$&$18.92\pm 0.82$&$16.16\pm 0.48$&$14.68\pm 0.31$\\
12:22:57.6618&15:49:25.174&$19.34\pm 0.68$&$18.85\pm 0.81$&$16.78\pm 0.52$&$15.17\pm 0.33$\\
12:22:57.4020&15:49:31.175&$17.93\pm 0.62$&$17.92\pm 0.78$&$15.46\pm 0.49$&$13.79\pm 0.31$\\
Filament 18, & red diamond & & & & \\
12:22:58.3898&15:50:57.799&$17.51\pm 1.02$&$17.22\pm 1.27$&$14.88\pm 0.64$&$13.21\pm 0.37$\\
12:22:57.8441&15:51: 4.924&$19.04\pm 1.22$&$18.54\pm 1.47$&$16.33\pm 0.73$&$14.65\pm 0.43$\\
12:22:57.5323&15:51: 7.924&$19.08\pm 1.23$&$18.84\pm 1.48$&$17.85\pm 0.86$&$17.71\pm 0.58$\\
Filament 19, & cyan {\it o} & & & & \\
12:22:59.1169&15:49:19.923&$18.53\pm 0.67$&$18.58\pm 0.83$&$16.58\pm 0.51$&$14.97\pm 0.32$\\
12:22:59.0390&15:49:28.173&$19.35\pm 0.69$&$19.36\pm 0.84$&$16.68\pm 0.48$&$15.12\pm 0.30$\\
12:22:58.7012&15:49:30.423&$17.27\pm 0.57$&$16.71\pm 0.65$&$15.18\pm 0.39$&$13.69\pm 0.25$\\
12:22:58.3634&15:49:35.299&$17.71\pm 0.62$&$17.73\pm 0.76$&$15.88\pm 0.45$&$14.22\pm 0.28$\\
12:22:58.1036&15:49:40.174&$18.06\pm 0.65$&$17.77\pm 0.79$&$15.90\pm 0.48$&$14.29\pm 0.31$\\
12:22:57.5320&15:49:49.174&$19.10\pm 0.71$&$19.04\pm 0.88$&$16.82\pm 0.53$&$15.00\pm 0.34$\\
Filament 20, & blue $+$ & & & & \\
12:22:58.5970&15:48:40.923&$18.14\pm 0.71$&$17.83\pm 0.84$&$17.38\pm 0.51$&$16.32\pm 0.34$\\
12:22:58.3372&15:48:41.299&$20.50\pm 0.76$&$22.50\pm 1.03$&$16.97\pm 0.47$&$14.80\pm 0.29$\\
12:22:58.1034&15:48:45.049&$19.27\pm 0.68$&$19.12\pm 0.82$&$18.55\pm 0.50$&$16.43\pm 0.31$\\
Filament 21, & magenta {\it o} & & & & \\
12:22:59.5585&15:48:54.797&$17.74\pm 0.49$&$17.44\pm 0.56$&$15.54\pm 0.27$&$14.18\pm 0.17$\\
12:22:59.0128&15:48:54.048&$17.92\pm 0.52$&$18.28\pm 0.64$&$16.08\pm 0.30$&$14.16\pm 0.18$\\
12:22:58.7790&15:48:55.923&$17.85\pm 0.48$&$17.41\pm 0.54$&$14.95\pm 0.25$&$13.25\pm 0.16$\\
12:22:58.5971&15:48:56.673&$17.60\pm 0.47$&$16.87\pm 0.51$&$15.63\pm 0.27$&$14.35\pm 0.18$\\
Filament 22, & blue diamond & & & & \\
12:23: 0.9365&15:50:45.045&$18.44\pm 0.95$&$18.27\pm 1.15$&$16.17\pm 0.55$&$14.48\pm 0.34$\\
12:23: 0.8066&15:50:49.545&$18.99\pm 1.02$&$18.68\pm 1.22$&$17.92\pm 0.67$&$15.90\pm 0.40$\\
Filament 23, & red {\it x}  & & & & \\
12:23: 1.3263&15:50:40.169&$17.31\pm 0.82$&$17.07\pm 0.96$&$14.86\pm 0.43$&$13.21\pm 0.26$\\
12:23: 1.3783&15:50:46.544&$18.94\pm 0.91$&$18.62\pm 1.06$&$16.86\pm 0.51$&$15.00\pm 0.31$\\
12:23: 1.5083&15:50:49.169&$18.27\pm 0.80$&$17.73\pm 0.91$&$16.28\pm 0.45$&$14.54\pm 0.28$\\
Filament 24, & red triangle & & & & \\
12:23: 1.4819&15:50: 0.419&$18.56\pm 0.97$&$18.45\pm 1.19$&$15.88\pm 0.67$&$14.07\pm 0.42$\\
12:23: 1.3520&15:50: 8.294&$18.52\pm 0.96$&$18.21\pm 1.17$&$16.30\pm 0.69$&$14.71\pm 0.44$\\
12:23: 1.4821&15:50:21.044&$19.38\pm 1.04$&$19.22\pm 1.28$&$16.51\pm 0.68$&$14.75\pm 0.41$\\
Filament 25, & blue {\it x}  & & & & \\
12:23: 2.4698&15:50:42.792&$18.74\pm 1.10$&$18.71\pm 1.38$&$16.51\pm 0.72$&$15.00\pm 0.44$\\
12:23: 2.5998&15:50:52.917&$17.93\pm 1.02$&$17.83\pm 1.26$&$15.67\pm 0.62$&$13.95\pm 0.37$\\
12:23: 2.5999&15:50:58.542&$20.25\pm 1.32$&$20.18\pm 1.62$&$19.27\pm 0.84$&$17.10\pm 0.51$\\
12:23: 2.4960&15:51: 3.792&$20.05\pm 1.35$&$20.23\pm 1.71$&$18.10\pm 0.80$&$16.27\pm 0.49$\\
12:23: 2.4181&15:51: 8.667&$18.58\pm 1.20$&$18.40\pm 1.47$&$16.70\pm 0.71$&$14.91\pm 0.43$\\
Filament 26, & blue triangle & & & & \\
12:23: 2.6245&15:48:48.417&$18.55\pm 0.88$&$18.29\pm 1.07$&$16.37\pm 0.56$&$14.95\pm 0.36$\\
12:23: 2.8324&15:48:50.666&$20.51\pm 1.04$&$19.70\pm 1.23$&$16.89\pm 0.63$&$15.29\pm 0.40$\\
12:23: 3.2222&15:48:54.790&$19.09\pm 0.98$&$19.05\pm 1.21$&$16.76\pm 0.66$&$15.30\pm 0.42$\\
Filament 27, & cyan $+$ & & & & \\
12:23: 3.7676&15:48:36.039&$18.95\pm 1.18$&$19.18\pm 1.50$&$16.75\pm 0.85$&$15.28\pm 0.55$\\
12:23: 4.0795&15:48:45.413&$18.13\pm 1.02$&$18.03\pm 1.25$&$16.66\pm 0.76$&$15.32\pm 0.50$\\
12:23: 4.3654&15:48:51.038&$18.72\pm 1.07$&$18.10\pm 1.25$&$16.15\pm 0.68$&$14.32\pm 0.42$\\
12:23: 4.5734&15:48:58.162&$19.99\pm 1.26$&$19.39\pm 1.49$&$17.06\pm 0.79$&$15.30\pm 0.49$\\
\enddata
%\tablenotetext{1}{In order of average RA}
%\tablenotetext{2}{in magnitudes}
\label{tab}
\end{deluxetable}
%\end{longtable}

\newpage

%fig1

\newpage

%fig2

\newpage

%fig3

%fig4

\newpage
%fig5

% note >3 because variance divided by n-1 and if there are 3 clumps, n=1 measure of the separation difference

%fig6

\newpage
%fig7

%fig8

%fig9

\end{document}